# Beyond Activity Space: Detecting Communities in Ecological Networks

Wenna Xi[1]

Catherine A. Calder[2]

Christopher R. Browning[3]

[1] Division of Biostatistics, College of Public Health, The Ohio State University

[2] Department of Statistics, The Ohio State University

[3] Department of Sociology, The Ohio State University

# Abstract

Emerging research suggests that the extent to which activity spaces – the collection of an individual's routine activity locations – overlap provides important information about the functioning of a city and its neighborhoods. To study patterns of overlapping activity spaces, we draw on the notion of an ecological network, a type of two-mode network with the two modes being individuals and the geographic locations where individuals perform routine activities. We describe a method for detecting "ecological communities" within these networks based on shared activity locations among individuals. Specifically, we identify latent activity pattern profiles, which, for each community, summarize its members' probability distribution of going to each location, and community assignment vectors, which, for each individual, summarize his/her probability distribution of belonging to each community. Using data from the Adolescent Health and Development in Context (AHDC) Study, we employ latent Dirichlet allocation (LDA) to identify activity pattern profiles and communities. We then explore differences across neighborhoods in the strength, and within-neighborhood consistency of community assignment. We hypothesize that these aspects of the neighborhood structure of ecological community membership capture meaningful dimensions of neighborhood functioning likely to co-vary with economic and racial composition. We discuss the implications of a focus on ecological communities for the conduct of "neighborhood effects" research more broadly.

# Introduction

Research on the effects of neighborhood contexts has a long history in social science and public health, but has increased exponentially in the last two decades in response to seminal research (Wilson 1987, 1996; Kawachi and Berkman 2003) and the increasing incorporation of data on neighborhood context into large-scale studies of the social context of health (Sampson, Raudenbush, and Earls 1997; Morenoff 2003; Sastry et al. 2006; Harris 2013). Although "neighborhood effects" research has produced compelling evidence of geographic context effects on a range of outcomes, the field has seen a recent shift away from exclusive focus on the neighborhood of residence (typically operationalized using administrative units such as census tracts, census block groups, or zip codes) as the critical non-home exposure context. Scholars from a range of disciplines have advocated for considering activity spaces, or the geographic locations an individual frequents as part of his or her daily routine, to more accurately capture the influence of social and spatial context (Golledge and Stimson 1997; Albert, Gesler, and Levergood 2000; Kwan 2009; Matthews and Yang 2013; Browning and Soller 2014). Activity space has been used to study non-residential segregation (Wong and Shaw 2011), depression (Vallée et al. 2011), health-related behaviors (Vallée et al. 2010; Zenk et al. 2011), and accessibility to healthcare opportunities (Sherman et al. 2005). Overviews of research employing the notion of activity space can be found in Ren (2016) and Perchoux et al. (2013).

A focus on the characteristics of potentially idiosyncratic activity spaces naturally leads to investigation of individual-level exposure-outcome links. For example, Zenk et al. (2011) studied the association between environmental characteristics of an individual's activity space (e.g., fast food outlet density, supermarket availability, park land use) and his/her dietary and physical activity behaviors. To the best of our knowledge, however, little is known about the implications of activity spaces in the aggregate for the functioning of neighborhoods and cities as a whole. Vallée et al. (2010, 2011) looked at the interaction effect of activity space and neighborhood via multilevel logistic regression models. They found that neighborhood environment is associated with health behaviors and outcomes of individuals with limited activity space (limited relative to their neighborhood); a limited activity space is protective for depression for individuals living in advantaged neighborhoods and a risk factor for those living in disadvantaged neighborhoods (Vallée et al. 2011).

In our previous work (Browning and Soller 2014; Browning et al. 2017b), we explored the effects of overlapping activity spaces through the notion of an *ecological (eco-) network.* An eco-network can be understood as a two-mode network with the modes being individuals residing within an urban environment and the geographic locations they visit. Network ties are defined only between nodes of different modes (i.e., individuals can only be connected to locations). Using data from the Los Angeles Family and Neighborhood Study (L.A.FANS) (Sastry et al. 2006), Browning et al. (2017b) estimated eco-networks at the neighborhood level, where residents are tied through shared activity locations (in and beyond the neighborhood boundary). They found that neighborhood eco-network intensity (the extent to which household dyads encounter one another through multiple locations) is positively associated with collective efficacy (Sampson, Raudenbush, and Earls 1997), intergenerational closure (Coleman 1990; Sampson, Morenoff, and Earls 1999), and social network interaction and exchange (Morenoff 2003; Browning, Dietz, and Feinberg 2004), and that neighborhood eco-network extensity (the average proportion of all households in the neighborhood to which a given household is tied through any routine activity location) is positively associated with collective efficacy and intergenerational closure.

A focus on features of neighborhood eco-networks emphasizes shared activity patterns among neighborhood residents, but neglects the larger structure of everyday intersection characterizing urban areas as a whole. Scaling the eco-network up to the city level provides an opportunity to identify clusters of individuals linked together by commonality in the set of routine locations they visit – or what we call *ecological (eco-) communities.* The eco-community concept acknowledges the importance of both direct activity space exposures as well as indirect ties to individuals and places that are socially (but not necessarily spatially) proximate through mobility flows. Extant research suggests that indirect ties to places may be consequential for neighborhood and individual-level wellbeing. Mears and Bhati (2006), for instance, found that the crime rates of compositionally homophilous neighborhoods are associated (regardless of proximity), inferring that mobility-based ties among these neighborhoods drive their correlated outcomes. In one of the few studies that empirically examines the consequences of indirect exposures, Graif, Lungeanu, and Yetter (2017) found evidence that the crime rates of places to which neighborhood residents commute are associated with focal neighborhood crime. These studies suggest that aggregate links in the eco-network have consequences for the experience of any given location, regardless of an individual's own exposures. The eco-community concept extends the logic of indirect influence to emphasize the potential for higher order, ramifying influence within clusters of interconnected individuals and places.

The utility of identifying eco-communities depends, in part, on the extent to which they capture contexts that are empirically distinct from residential neighborhoods. The conventional idealized image of the neighborhood tends to be conflated with community, with residents presumed to encounter one another on a regular basis in the course of routines such as grocery shopping, school attendance, religious events, and leisure activities. Evidence of cross-neighborhood variation in the extent of shared activities (Browning et al. 2017b) suggests the potential for patterns of community membership to vary substantially across neighborhoods as well. The extent of such variability is potentially significant for understanding neighborhood functioning. Neighborhoods in which residents are tied to a single community are likely to experience higher levels of social cohesion and be better equipped to take action on behalf of shared goals. In contrast, neighborhoods where residents are tied to a variety of distinct communities may experience greater fragmentation and lack the capacity for conjoint pro-social action.

Consistent with prior research on the wide-ranging consequences of neighborhood inequality (Sampson 2012), we hypothesize that the neighborhood structure of community membership is organized by economic disadvantage and segregation. Poor and segregated neighborhoods may be characterized by deinstitutionalization (Wilson 1987, 1996; Small 2006; Small and McDermott 2006) with limited access to jobs, schools, commerce, and other amenities. In turn, residents may be forced to seek resources outside the neighborhood, leading to potentially idiosyncratic activity space exposures and community membership patterns. High levels of fear and mistrust may also lead residents of disadvantaged neighborhoods to avoid local activity locations, also contributing to spatial dispersion in activity patterns and fragmented community ties. Avoidance associated with fear may also lead to more time spent at home, resulting in fewer routine activities overall and weaker community embeddedness. Relatedly, residential instability (with which poverty and segregation are associated) may contribute to peripheral membership in multiple communities as individuals mix routine activities based on a prior location or job with new destinations. Even when individuals residing in disadvantaged neighborhoods are closely tied to communities (structurally), however, the set of mobility constraints these individuals face is likely to result in a weaker association between the strength of community attachments and the consistency of community membership across neighbors. In short, we expect more advantaged, less segregated neighborhoods to be characterized by a higher likelihood of consistent or homogenous community membership, stronger structural attachments to the communities with which they are affiliated, and a stronger association between community attachment and consistency of community membership.

We draw on both the eco-network and eco-community concepts to study the patterns of overlap in individuals' activity spaces within a larger urban environment. First, we describe a methodology for the identification of ecological communities. Our approach relies on latent Dirichlet allocation (LDA) in order to generate *activity pattern profiles,* or probability distributions of visiting each location at the community level, and *community assignment vectors* for each individual. These vectors provide the basis upon which to assign *modal* community membership as well as determine the *strength* of community attachments, indicated by high probability of membership in single community and low probabilities of alternative community membership. At the neighborhood level, investigation of variability in the distribution of community membership probabilities across individuals captures the *consistency* of community ties within neighborhoods. Second, we apply this approach to an observed eco-network for the Columbus, Ohio area using activity space data on 1,307 caregivers from the Adolescent Health and Development in Context (AHDC) Study (Browning et al. 2016). Third, we extract and investigate measures of the strength and consistency of community membership at the neighborhood level. Finally, in an illustrative analysis, we examine the association between neighborhood level economic and racial composition and the consistency of community membership, testing the hypothesis that neighborhood disadvantage and segregation tend to fragment community membership within neighborhoods.

## Data

**The AHDC Study**

The AHDC Study is a longitudinal survey that focuses on the impact of social and spatial environments on the health and development of urban youth (Browning et al. 2016, 2017a). The study involved adolescents between eleven and seventeen years old and their caregivers residing within the I-270 outerbelt in the Columbus, Ohio metropolitan area. The first wave of the AHDC data was collected between April 2014 and July 2016, and involved three data collection periods: the entrance survey, the ecological momentary assessment (EMA) week, and the exit survey. As part of the entrance survey, caregivers of the focal youth were asked about the places they go during a typical week, including weekend, with the following location types: workplace (main job), school/college/training program, library, church or other religious place, grocery store, relative's house, friend's house, rec center/park/sport's facility, restaurant, store or other business, civic organization, neighborhood organization (block watch, etc.), and other. The locations reported could be outside of the I-270 outerbelt. For each location type, multiple locations could be provided. The same location could be provided as responses to multiple location types as well. The latitude and longitude coordinates of each location were collected via the Google Maps API. The home address and reported locations of each caregiver were then aggregated to the census block group. An ecological network can be created, represented as a matrix, with rows indexing caregivers, columns indexing block groups, and entries being the number of times caregivers reported a block group as containing the location of a routine activity.

Among the 1,404 caregivers who were interviewed as part of the AHDC Study, five of them did not share any visited block groups with others, one individual did not provide any activity locations, and one individual provided a home address outside of the I-270 outerbelt. These seven caregivers are excluded from the following analyses. In order to study variability in activity pattern profiles at the neighborhood level, sufficient density of sampled caregivers within a neighborhood is needed. Therefore, an additional ninety caregivers whose home locations are in census tracts that contain fewer than four (<4) caregivers are excluded. The 1,307 caregivers included in the study reported activity locations in a total of 883 block groups.

To protect the identities of the study participants, home addresses are not allowed to be released according to the study protocol. Therefore, a random point from the same block group as the home address is used to represent the home location in all maps presented in this paper.

**The AHDC Caregivers' Ecological Network**

Columbus, Ohio is the capital of the state of Ohio and the 14-th most populous city in the United States. It is often described as an "average" American city and has historically served as test market city for the vetting of new products (Holloway et al. 1999). Columbus is also a very segregated city, both racially and economically (Florida and Mellander 2015).

Figure 1A provides a brief summary of the study area in Columbus (inside I-270). The black shading in the background indicates the proportion of black residents at the block group level, with darker shading indicating a higher proportion of blacks. The data come from the 2009-2013 American Community Survey 5-year (ACS 5-yr) released by the U.S. Census Bureau. The Hispanic and Asian populations in Columbus are relatively small (5.6 percent and 4.1 percent, respectively, based on the 2010 census population); therefore, a higher percentage of black residents often indicates a lower percentage of white residents. The west side of Columbus is mostly white; blacks are concentrated in the east side of the city, with the exception of Bexley, an affluent suburb in the southeast of Columbus. Except for Bexley, all other wealthy suburbs (inside I-270) are in the west side of the city, including Grandview Heights, Upper Arlington, and Dublin. While Upper Arlington, Dublin, and Bexley mostly consist of families with kids, Grandview Heights attracts more childless young professionals. Easton Town Center, one of the premier shopping destinations in the Columbus metropolitan area, locates in the northeast of Columbus. The Central Business District (CBD) of Columbus is in downtown, which contains numerous businesses, organizations, educational institutions, and entertainment options.

Figure 1B summarizes caregivers' home locations and activity locations in the study region (inside I-270). Both home and activity locations shown here are approximations to the actual addresses. In total, the 1,307 caregivers provide 7,092 routine activity locations that are inside I-270.

Figure 2 summarizes the number of activity locations caregivers report in each block group. Figure 2A summarizes the block groups by the total number of routine activity locations among all caregivers, with darker shading indicating higher visit frequencies. Caregivers with different characteristics are attracted to different block groups; for illustrative purposes, three block groups with interesting caregiver patterns are summarized below.

Figure 2B shows the approximate location of residence for caregivers who regularly visit the most popular block group, the one in the northeast that contains Easton Town Center. Caregivers who regularly visit this block group live in different areas of the city, but are mainly from the east side. The majority of those who live in the northeast are black (results not shown). Caregivers who regularly visit this block group have a wide range of household incomes (results not shown). Compared to the west side, a larger proportion of caregivers living in the northeast visit this block group.

Figure 2C shows the approximate location of residence for caregivers who regularly visit the fifth most popular block group, located in Bexley, Ohio. This block group contains Bexley Middle School and Bexley High School. All caregivers who regularly visit this block group live in Bexley. This is a group who are mostly white, with dispersed household incomes (results not shown). A very large proportion of caregivers living in Bexley visit this block group, while none of the caregivers living outside of Bexley visit it on a regular basis.

Figure 2D shows the approximate location of residence for caregivers who regularly visit the tenth most popular block group, located in downtown Columbus. Although most of them live on the east side of the city, their home locations are very dispersed. They are mostly black, especially those from the east side, and most of them are middle class (annual household income between $30K and $150K; results not shown). The proportion of caregivers visiting this block group is very homogeneous across the city; however, a few block groups close to downtown Columbus have slightly higher proportion visiting than the rest.

# Methods

Despite the abundance of existing methods for one-mode network analysis, methods for two-mode network analysis still remain underdeveloped (Wasserman and Faust 1994; Field et al. 2006). Two-mode networks can be visualized using the biplot. Some existing statistical methods for analyzing two-mode networks include Galois Lattices (Wasserman and Faust 1994) and correspondence analysis (CA) (Borgatti, Everett, and Johnson 2013). Some existing statistical methods for detecting communities in two-mode networks include the extended $p^*$ models (Field et al. 2006), the dual-projection approach (Melamed 2014), and the bilinear mixed-effects model (Jia, Calder, and Browning 2014; Jia 2016). Most of these community detection methods, however, only assign one community to each individual and do not consider the strength of the community assignment. As an alternative to these methods, we propose the use of latent Dirichlet allocation (LDA) (Blei, Ng, and Jordan 2003), which allows us to study the community structure by assigning a probability distribution over locations for each community. Although originally developed for and mainly used in text data mining, LDA has previously been used to study the patterns in the locations of individuals' check-ins on Facebook (Chang and Sun 2011). As Chang and Sun (2011) pointed out, LDA "has the potential to find culturally similar 'neighborhoods' even if they span different areas of a city." Following this suggestion, we use LDA to study individuals' activity patterns in the aggregate and detect communities in ecological networks.

Previous studies have shown that non-negative matrix factorization (NMF) with Kullback-Leibler (KL) divergence is very similar to LDA (Ding, Li, and Peng 2008; Paisley, Blei, and Jordan 2014). However, NMF does not have the Dirichlet constraints as in LDA, which allow the probabilistic interpretation of our activity pattern profiles and community assignment vectors needed for interpreting our findings.

Using the caregivers' data from the AHDC Study, we first employ LDA to identify *activity pattern profiles* which are probability distributions over a discrete set of locations, which in our analysis of the AHDC data are census block groups. These activity pattern profiles are what define *ecological (eco-) communities*. In turn, our method finds, for each individual, a *community assignment vector*, which is a discrete probability distribution over the eco-communities (which corresponds to the activity pattern profiles) and we define the community to which an individual belongs with the highest probability to be the individual's *modal community*. After identifying activity pattern profiles and modal communities, we further investigate the spatial structure of our findings by looking at, for example, the extent to which modal community assignment is spatially clustered when individuals are geographically positioned by their home locations. We then examine the consistency of community assignment by employing the *Gini coefficient* to look at the strength of community assignment for each caregiver. We aggregate the Gini to the neighborhood level to measure the average community assignment strength. Finally, we use *total variation* from the Aitchison geometry to examine the variability in patterns of community assignment across individuals to measure the consistency of community assignment at the neighborhood level.

**Latent Dirichlet Allocation (LDA) for Ecological Networks**

Text data mining, as a multidisciplinary field involving statistics, machine learning, and data mining, is the process of extracting patterns from unstructured text documents (Tan 1999). In text data mining, latent Dirichlet allocation (LDA) (Blei, Ng, and Jordan 2003) is often viewed as a method to discover the hidden "topics" from documents in a corpus. LDA assumes the existence of a latent variable, called a "topic," such that each document consists of a random mixture of latent topics, and each topic is a distribution over words in the corpus. Although the settings differ, detecting communities in ecological networks resembles the task of identifying topics in a corpus. Similar to having documents and words and assuming a latent topic structure in a corpus, in ecological networks, we have individuals and locations and can assume a latent community structure, which are defined by activity pattern profiles.

To specify the model, we let $Y_{n_{ij}}$ be the indicator that the $n_i$-th location individual $i$ visits is location $j$, where $n_i = 1, \ldots, N_i, i = 1, \ldots, I, j = 1, \ldots, J$. LDA assumes

$$Y_{n_{ij}} | p_{n_{ij}} \overset{iid}{\sim} Bernoulli(p_{n_{ij}}),$$

where $p_{n_{ij}} \overset{def}{=} \mathbf{W}_{i\cdot} \times \mathbf{H}_{\cdot j}$, $\mathbf{W}$ is an $I \times K$ matrix with $W_{ik}$ being the probability of individual $i$ belonging to community $k$, and $\mathbf{H}$ is a $K \times J$ matrix with $H_{kj}$ being the probability of an individual from community $k$ visiting location $j$, $k = 1, \ldots, K$. In addition, LDA assumes the priors of $\mathbf{W}$ and $\mathbf{H}$ are

$$\mathbf{W}_{i\cdot} \overset{iid}{\sim} Dirichlet(\boldsymbol{\alpha})$$

and

$$\mathbf{H}_{k\cdot} \overset{iid}{\sim} Dirichlet(\boldsymbol{\beta}),$$

where $\boldsymbol{\alpha}$ and $\boldsymbol{\beta}$ are known constant vectors of dimension $K$ and $J$, respectively.

The Dirichlet distribution imposes the following constraints on $\mathbf{W}_{i\cdot}$ and $\mathbf{H}_{k\cdot}$ to ensure that they are valid (discrete) probability distributions:

1. $\sum_{k=1}^{K} W_{ik} = 1$ for all $i = 1, \ldots I$, and

2. $\sum_{k=1}^{K} H_{kj} = 1$ for all $k = 1, \ldots K$.

We call $\mathbf{W}_{i\cdot} = (W_{i1}, \ldots, W_{iK})$, the probability distribution over communities for individual $i$, the *community assignment vector* of individual $i$; $\mathbf{H}_{k\cdot} = (H_{k1}, \ldots, H_{kJ})$, the probability distribution over locations for community $k$, the *activity pattern profile* of community $k$; and the community with the largest probability, $\mathrm{argmax}_{k}\{W_{ik}: k = 1, \ldots, K\}$, the *modal community* of individual $i$.

Let $l_{n_i}$ be the $n_i$-th location individual $i$ visits, and $c_i$ be the community picked for individual $i$ based on his/her community assignment vector, $n_i = 1, \ldots, N_i$, $i = 1, \ldots, I$. Then, for the $n_i$-th location, by the law of total probability, the probability of individual $i$ visiting location $j$ is $\sum_{k=1}^{K} P(c_i = k) \times P(l_{n_i} = j | c_i = k)$, where $P(c_i = k)$, the probability that individual $i$ belongs to community $k$, is equal to $W_{ik}$, and $P(l_{n_i} = j | c_i = k)$, the probability that individuals in community $k$ visit location $j$, is equal to $H_{kj}$. The LDA model can thus be understood via the following equation:

$$P(Y_{n_{ij}} = 1) = p_{n_{ij}} = P(l_{n_i} = j)$$

$$= \sum_{k=1}^{K} P(c_i = k) \times P(l_{n_i} = j | c_i = k)$$

$$= \sum_{k=1}^{K} W_{ik} \times H_{kj}$$

$$= \mathbf{W}_{i\cdot} \times \mathbf{H}_{\cdot j}.$$

We fit the LDA models using the 'LDA' function with the 'Gibbs' sampling method (Phan, Nguyen, and Horiguchi 2008) in the R package 'topicmodels' (Hornik and Grün 2011).

**Selecting the Number of Communities**

One assumption of LDA is that the number of communities is known. To find the optimal number of communities, we follow Blei, Ng, and Jordan (2003) and fit multiple models with different numbers of communities and compare their *perplexities*. In natural language processing, perplexity is used to evaluate the goodness-of-fit of the model (Azzopardi, Girolami, and van Risjbergen 2003; Blei, Ng, and Jordan 2003). In the context of ecological networks, perplexity is the exponential of the average negative log-likelihood for each individual (Bengio et al. 2003); therefore, smaller perplexity indicates better fit of the model (Blei, Ng, and Jordan 2003).

In detail, we first split our data (by individual) into a training set (90 percent) and a testing set (10 percent). For each number of communities ($K = 5$ to 140), we fit the LDA model on the training set, and calculate the perplexity on the validation set. The perplexity of a testing set with $I'$ individuals is defined as

$$Perplexity(Testing\ Set) = exp\{-\frac{\sum_{i=1}^{I'} \sum_{n_i=1}^{N_i} log\ p(l_{n_i})}{\sum_{i=1}^{I'} N_i}\}.$$

This process is repeated 20 times (20 different test/training sets) and the $K$ of the model with the smallest average (across test/training sets) perplexity is chosen. We compare the models using the 'perplexity' function in the R package 'topicmodels.'

# Results

**Community Detection**

Our cross-validation exercise indicates that the LDA model with 18 communities has the smallest average perplexity (average perplexity = 328.054). Therefore, for the analysis, the model with 18 communities is fitted to the entire data set ($I = 1,307, J = 902$). The estimated community assignment vector for caregiver $i$ is the posterior mean $\widehat{\mathbf{W}}_{i\cdot}$, the estimated activity pattern profile for community $k$ is the posterior mean $\widehat{\mathbf{H}}_{k\cdot}$, and the estimated modal community of caregiver $i$ is the one to which he/she has the highest probability of belonging ($\operatorname*{argmax}_{k}\{\widehat{W}_{ik}: k = 1,\ldots,18\}$), $i = 1,\ldots,1307$, $k = 1,\ldots,18$.

The result of LDA with eighteen pre-specified communities is shown in Figure 3. Here, caregivers are represented by their approximate home locations, which are colored according to their modal communities. Even though only caregivers' activity locations (not including their home locations) are used in the LDA analysis, clear spatial clusters of communities can still be observed. Therefore, individuals living close to each other have similar activity patterns.

Table 1 summarizes the sizes of the communities; the largest community consists of 118 caregivers (community 1), and the smallest 27 (community 18). Figure 4 summarizes the distributions of annual household income and race for each community. Although we do observe some extreme communities (e.g., caregivers in community 10 are mostly white and wealthy, and caregivers in community 18 are mostly black and about half of them have annual household income less than or equal to $30K), overall, compared to block-group level summaries (Figure 1A), distributions of race and household income are more mixed within each community.

Since the overlap of communities makes it difficult to see the spatial patterning of each individual community, separate figures of each community are shown in Appendix 2. For illustrative purposes, separate figures of communities 10, 12, 13, and 15 are shown in Figure 5. In each figure, points indicate the approximate home locations of caregivers in that community, with the shade indicating the probability of them belonging to it. Block groups are shaded to indicate the probability of caregivers from that community visiting it for a routine activity.

Community 10, as shown in Figure 5A, consists mainly of caregivers living in Upper Arlington. Interestingly, LDA separates caregivers in Upper Arlington from caregivers in its neighboring city Grandview Heights (see community 16 in Appendix 2). Community 13, as shown in Figure 5C, consists mostly of caregivers living in Bexley. Communities 10 and 13 (as well as community 16) are very compact, and consist mostly of caregivers who reside in or very close to that region. The most dominating block group of those two communities, though covered by the densely plotted home locations, locate inside the two suburbs, respectively.

Community 12, as shown in Figure 5B, is formed mostly by caregivers who frequent the tenth most visited block groups, the one located in downtown Columbus. Community 15, as shown in Figure 5D, is formed mostly by caregivers who frequent the most visited block group, the one that contains Easton Town Center. Caregivers in communities 12 and 15 have home locations more dispersed across the city, have lower household income (Figure 4A), and are racially more diverse (Figure 4B). As a general pattern, wealthier communities tend to be more geographically compact and whiter.

**Quality of Community Identification**

To demonstrate that the communities detected by LDA are effectively clustering caregivers with similar activity patterns, we estimate (via simulation from the fitted model) the probability that two caregivers share a routine activity location on whether they are assigned to the same modal community or not and as a function of the number of routine activity locations they report. The results of the simulation are shown in Figure 6. As a reference, the blue curve is the exact (calculated analytically) probability of two random caregivers (not necessarily having the same modal community) each visiting $n$ ($n = 1, ..., 35$) locations sharing at least one of the same locations visited (out of 883). Compared to caregivers from different communities, caregivers from the same community are more likely to have shared locations. When two caregivers from the same community each visit 10 out of 883 locations, the average probability of them sharing at least one routine activity location is 0.934 (SD = 0.059). When two caregivers from different communities each visit 10 out of 883 locations, the average probability of them sharing at least one routine activity location is 0.119 (SD = 0.070).

**Communities versus Neighborhoods**

Following the convention, census tracts are used as proxies for (residential) neighborhoods (Morland et al. 2002; Sampson 2019). In total, there are 201 census tracts inside (or partially inside) I-270, and the 1,307 caregivers reside in 138 of them.

On average, 28.725 percent (variance = 0.070) of the caregivers from the same neighborhood share the same modal community (distribution shown in Figure 7A; more numerical summaries in Table 2), and the number of different modal communities of caregivers from the same neighborhood is 4.594 (variance = 4.039; more numerical summaries in Table 3). Since the number of modal communities represented may be larger for neighborhoods with more caregivers, Figure 7B summarizes the number of communities by the number of caregivers in the neighborhood. The blue line is the regression line of the number of caregivers on the number of modal communities. Contrary to our expectation, the number of communities is negatively affected by the number of caregivers in the neighborhood (coefficient = -0.066, $p = 0.026$).

**Caregivers' Community Assignment Patterns**

To measure the strength of caregiver's community assignment, the Gini coefficient (Gini 1912) is calculated. The Gini coefficient, most commonly used in economics to quantify income or wealth inequality (Atkinson 1970), is a measure of statistical dispersion (Aquino, de Oliveira, and Barreto 2009). The Gini coefficient of caregiver $i$'s community assignment vector $\mathbf{W}_{i\cdot} = (W_{i1}, \ldots, W_{iK})$ is

$$\hat{G} = \frac{\sum_{k=1}^{K}\sum_{l=1}^{K}|\widehat{W}_{ik}-\widehat{W}_{il}|}{2K\sum_{k=1}^{K}\widehat{W}_{ik}},$$

where $\widehat{W}_{ik}$ is the estimated probability of caregiver $i$ belonging to community $k$, $k = 1, \ldots, K$ (Pyatt 1976). The Gini coefficient measures the strength of caregiver's attachment to the communities. If someone has equal probability of belonging to all communities, then his/her Gini coefficient is 0. If, in the other extreme, someone belongs to one community with probability 1 and other communities with probability 0, then his/her Gini coefficient is $1 - \frac{1}{K}$. The Gini coefficient is strongly correlated with the number of routine activity locations caregivers report ($R^2 = 0.883$, $p < 0.001$). This is not surprising because LDA detects communities based on the number of routine activity locations caregivers provide. As the number of locations a caregiver provides increases, more information is collected, and therefore LDA can assign him/her to communities with more confidence (higher probability). Controlling for the number of locations provided, the Gini coefficient is strongly associated with race ($p < 0.001$) and annual household income ($p < 0.001$), respectively. The Gini coefficient is higher among whites than blacks, and increases as the annual household income increases.

To understand the variation of caregivers' community assignment patterns at the neighborhood level, the mean Gini coefficient of all caregivers from the same neighborhood is calculated. The mean Gini is strongly associated with the mean number of locations provided by the caregivers from the neighborhood ($p < 0.001$), but is not associated with the number of caregivers in the neighborhood ($p = 0.270$) (results not shown). Controlling for the mean number of locations provided and the number of caregivers, the mean Gini coefficient is negatively associated with concentrated disadvantage ($p = 0.001$; see Table 4) and percent black ($p < 0.001$; see Table 5), respectively. The mean Gini is higher in advantaged and/or white neighborhoods and lower in disadvantaged and/or black neighborhoods. These findings are consistent with the expectation that neighborhood disadvantage and segregation are associated with weaker community attachment, on average.

It is possible, however, that everyone from the same neighborhood has a very high probability of belonging to one community, but they all belong to different communities. Therefore, in addition to the mean Gini coefficient, we estimate the neighborhood-level *consistency* of community assignment, measured by the total variation (Hron and Kubáček 2011) of the caregivers' community assignment vectors for each neighborhood. Since each caregiver's community assignment vector lies in a simplex, we use the total variation measure, as defined in the Aitchison geometry (Aitchison and Egozcue 2005). Let $\boldsymbol{X} = (X_1, \dots, X_K)^T$ be a random composition (a random vector) in the Aitchison geometry, where $X_k$ is the $k$-th random component (a random variable) with constraints $\sum_{k=1}^{K} X_k = 1$ and $X_k > 0$, $k = 1, \ldots, K$. For neighborhood $c$ with $n_c$ sampled caregivers, $\mathbf{W}_{c_1\cdot}^T, \ldots, \mathbf{W}_{c_{n_c}\cdot}^T$ are a random sample of $\mathbf{X}$, and correspondingly, $W_{c_1 k}, \ldots, W_{c_{n_c} k}$ are a random sample of $X_k$, $k = 1, \ldots, K$. We use the sample total variation of $\mathbf{W}_c^T = (\mathbf{W}_{c_1\cdot}^T, \ldots, \mathbf{W}_{c_{n_c}\cdot}^T)$ to estimate the total variation of neighborhood $c$'s community assignment:

$$\widehat{totvar}(\mathbf{W}_c^T) = \frac{1}{2K} \sum_{i=1}^{K} \sum_{j=1}^{K} \widehat{var}\left(\log \frac{X_i}{X_j}\right),$$

where $log$ is the natural logarithm, and $\widehat{var}\left(\log \frac{X_i}{X_j}\right) = \frac{1}{n_c - 1} \sum_{l=1}^{n_c} \left(\log \frac{\widehat{W}_{c_l i}}{\widehat{W}_{c_l j}} - \overline{\log \frac{\widehat{W}_{\cdot i}}{\widehat{W}_{\cdot j}}}\right)^2$ is the sample variance of $\log \frac{X_i}{X_j}$ for community $c$, $i = 1, \ldots K$, $j = 1, \ldots K$ (Hron and Kubáček 2011).

The total variation is strongly associated with the mean number of locations provided by the caregivers from the neighborhood ($p < 0.001$) and the number of caregivers in the neighborhood ($p = 0.014$), respectively (results not shown). Controlling for the mean number of locations provided and the number of caregivers, the total variation is negatively associated with mean Gini ($p < 0.001$) and positively associated with percent black ($p = 0.002$), respectively (results not shown). However, there is an interaction effect between mean Gini and percent black ($p = 0.005$; see Table 5). When percent black is low, the effect of mean Gini on total variation is negative. As percent black increases, the effect of mean Gini disappears (see Figure 8). In very white neighborhoods, the higher the mean Gini, the lower the total variation, whereas in very black neighborhoods, total variation is not associated with mean Gini.

## Discussion

In this paper, we use LDA to study the overlap in individuals' activity spaces in order to detect ecological communities and to identify activity pattern profiles of communities. Compared to NMF with KL divergence, the Dirichlet constraints in LDA provide us with probability distributions of individuals on communities and communities on locations, which are needed in our case to assign modal communities for individuals and to define activity pattern profiles for the communities. Compared to most of the existing methods to detect communities in two-mode networks (Field et al. 2006; Melamed 2014), the probability distributions of individuals on communities obtained from LDA allow us to further study the strength of individuals' community attachment and the consistency of their community membership at the neighborhood level.

We illustrate our methods with the caregivers' data from the AHDC Study, and further explore the strength of caregivers' community attachment and the consistency of their community membership at the neighborhood level. Using the perplexity of LDA models as the criterion, 18 communities are detected among the 1,307 caregivers residing inside the I-270 outerbelt of Columbus, Ohio. At the neighborhood level, we find that the strength of caregivers' community attachment is associated with concentrated disadvantage and percent black, respectively, controlling for the mean number of locations caregivers provide and the number of caregivers residing in the neighborhood. We also find that the consistency of caregivers' community membership is associated with the interaction between the strength of community attachment and the percentage of black residents, controlling for the mean number of locations they provide and the number of caregivers residing in the neighborhood: Among white neighborhoods, the stronger the residents are attached to one community, the higher the likelihood that they all belong to one community, whereas among predominantly black neighborhoods, the consistency of residents' community membership is not associated with the strength of their community attachment.

Our conclusion is robust to the number of communities in the LDA model; the LDA model with 25 communities also identifies caregivers' in Bexley, Grandview Heights, and Upper Arlington as separate communities and leads to the same conclusions about the strength and consistency of community attachment.

The implications of these findings for understanding urban neighborhood functioning are potentially significant. Fundamentally, acknowledgement of the wide-ranging, extra-neighborhood exposures experienced by urban residents is largely absent from the extant neighborhood literature. Our analysis sheds light not only on the prevalence of these exposures, but clustering of shared routines that operates independently of residential neighbor proximity. The basis of ecological communities in actual patterns of shared exposure suggests the possibility that characteristics of these units may contribute to individual-level health and wellbeing outcomes, above and beyond the influence of residential neighborhoods or individual level activity space exposures.

Second, neighborhood research has largely focused on internally-oriented characteristics of residential neighborhoods such as structural disadvantage and social network ties among neighbors to capture the conditions that promote or inhibit neighborhood social cohesion. By mapping patterns of community attachment and membership consistency onto the neighborhood, our approach offers the potential for identifying previously neglected sources of neighborhood cohesion and fragmentation. Wealthier, whiter neighborhoods are more likely to share a common community and to experience stronger associations between community attachment and community consistency. Future research may benefit from efforts to understand how structurally advantaged neighborhoods tap into shared communities to enhance neighborhood wellbeing.

Finally, the ecological network and community concepts point to more sophisticated research questions regarding the embeddedness of residents and neighborhoods in larger urban systems of interaction. What kinds of urban arrangements – physical and social – tend to promote more diverse or homogeneous ecological communities? What implications do these patterns hold for variability in the experience of cohesion across cities? Although a promising direction for urban research, availability of data resources capturing shared routines across urban residents remains limited. A recent exception is the work leveraging geo-referenced social media to proxy exposures (Wang et al. 2018). Yet these data are limited by uncertainty regarding the representativeness of participating individuals and geo-tagged locations.

In terms of limitations of our study, the AHDC caregiver reports include in some cases multiple locations for each location type and caregivers can report the same location for multiple types of routine activities. Therefore, the counts we model are not the number of times a caregiver visits a block group. Rather, they indicate the number of different activity types a caregiver conducts at a certain block group. In future research, we plan to explore different definitions of ecological communities that account for differences in the "importance" of different locations in an individual's activity space (e.g., by taking into account the time spent or the frequency of visits). These and other refinements will enhance the quality of eco-community estimates – a new and potentially significant contextual unit for understanding the substantial variability in outcomes across urban residents.

# References

Aitchison, J., and J. J. Egozcue. 2005. Compositional data analysis: Where are we and where should we be heading? *Mathematical Geology* 37 (7):829–50.

Albert, D. P., W. M. Gesler, and B. Levergood. 2000. *Spatial analysis, GIS and remote sensing: Applications in the health sciences.* Chelsea, MI: Ann Arbor Press.

Aquino, R., N. F. de Oliveira, and M. L. Barreto. 2009. Impact of the family health program on infant mortality in Brazilian municipalities. *American Journal of Public Health* 99 (1):87–93.

Atkinson, A. B. 1970. On the measurement of inequality. *Journal of Economic Theory* 2 (3):244–63.

Azzopardi, L., M. Girolami, and K. van Risjbergen. 2003. Investigating the relationship between language model perplexity and IR precision-recall measures. In *Proceedings of the 26th annual international ACM SIGIR conference on research and development in information retrieval*, 369–70. New York, NY: ACM Press. http://doi.acm.org/10.1145/860435.860505.

Bengio, Y., R. Ducharme, P. Vincent, and C. Jauvin. 2003. A neural probabilistic language model. *Journal of Machine Learning Research* 3 (Feb):1137–55.

Blei, D. M., A. Y. Ng, and M. I. Jordan. 2003. Latent Dirichlet allocation. *Journal of Machine Learning Research* 3 (Jan):993–1022.

Borgatti, S. P., M. G. Everett, and J. C. Johnson. 2013. *Analyzing social networks*. London: SAGE Publications Limited.

Browning, C. R., C. Calder, B. Soller, A. L. Smith, and B. Boettner. 2016. Measuring collective efficacy using georeferenced location reports: The Adolescent Health and Development in Context Study. Paper presented at Annual meetings of the American Society of Criminology, New Orleans, LA, November 16–19.

Browning, C. R., C. A. Calder, J. L. Ford, B. Boettner, A. L. Smith, and D. Haynie. 2017a. Understanding racial differences in exposure to violent areas: Integrating survey, smartphone, and administrative data resources. *The ANNALS of the American Academy of Political and Social Science* 669 (1):41–62.

Browning, C. R., C. A. Calder, B. Soller, A. L. Jackson, and J. Dirlam. 2017b. Ecological networks and neighborhood social organization. *American Journal of Sociology* 122 (6):1939–88.

Browning, C. R., R. D. Dietz, and S. L. Feinberg. 2004. The paradox of social organization: Networks, collective efficacy, and violent crime in urban neighborhoods. *Social Forces* 83 (2):503–34.

Browning, C. R., and B. Soller. 2014. Moving beyond neighborhood: Activity spaces and ecological networks as contexts for youth development. *Cityscape* 16 (1):165–96.

Chang, J., and E. Sun. 2011. Location3: How users share and respond to location-based data on social networking sites. In *Proceedings of the fifth international AAAI conference on weblogs and social media*, 74–80. Menlo Park, CA: AAAI Press.

Coleman, J. S. 1990. *Foundations of social theory*. Cambridge, MA: Harvard University Press.

Ding, C., T. Li, and W. Peng. 2008. On the equivalence between non-negative matrix factorization and probabilistic latent semantic indexing. *Computational Statistics & Data Analysis* 52 (8):3913–27.

Field, S., K. A. Frank, K. Schiller, C. Riegle-Crumb, and C. Muller. 2006. Identifying positions from affiliation networks: Preserving the duality of people and events. *Social Networks* 28 (2):97–123.

Florida, R., and C. Mellander. 2015. Segregated city: The geography of economic segregation in America's metros. Martin Prosperity Institute, Toronto.

Gini, C. 1912. *Variabilità e mutabilità: Contributo allo studio delle distribuzioni e delle relazioni statistiche* (*Variability and mutability: Contribution to the study of distributions and statistical reports*). Bologna: Tipografia di Paolo Cuppini.

Golledge, R. G., and R. J. Stimson. 1997. *Spatial behavior: A geographic perspective.* New York, NY: Guilford Press.

Graif, C., A. Lungeanu, and A. M. Yetter. 2017. Neighborhood isolation in Chicago: Violent crime effects on structural isolation and homophily in inter-neighborhood commuting networks. *Social Networks* 51 :40–59.

Harris, K. M. 2013. The Add Health study: Design and accomplishments. Carolina Population Center, University of North Carolina at Chapel Hill, Chapel Hill, NC.

Holloway, S. R., D. Bryan, R. Chabot, D. M. Rogers, and J. Rulli. 1999. Race, scale, and the concentration of poverty in Columbus, Ohio, 1980 to 1990. *Urban Geography* 20 (6):534–51.

Hornik, K., and B. Grün. 2011. topicmodels: An R package for fitting topic models. *Journal of Statistical Software* 40 (13):1–30.

Hron, K., and L. Kubáček. 2011. Statistical properties of the total variation estimator for compositional data. *Metrika* 74 (2):221–30.

Jia, Y. 2016. Generalized bilinear mixed-effects models for multi-indexed multivariate data. PhD diss., The Ohio State University.

Jia, Y., C. A. Calder, and C. R. Browning. 2014. Bilinear mixed-effects models for affiliation networks. arXiv preprint arXiv:1406.5954.

Kawachi, I., and L. F. Berkman. 2003. *Neighborhoods and health*. New York, NY: Oxford University Press.

Kwan, M.-P. 2009. From place-based to people-based exposure measures. *Social Science & Medicine* 69 (9):1311–13.

Matthews, S. A., and T.-C. Yang. 2013. Spatial polygamy and contextual exposures (spaces) promoting activity space approaches in research on place and health. *American Behavioral Scientist* 57 (8):1057–81.

Mears, D. P., and A. S. Bhati. 2006. No community is an island: The effects of resource deprivation on urban violence in spatially and socially proximate communities. *Criminology* 44 (3):509–48.

Melamed, D. 2014. Community structures in bipartite networks: A dual-projection approach. *PLoS ONE* 9 (5):e97823.

Morenoff, J. D. 2003. Neighborhood mechanisms and the spatial dynamics of birth weight. *American Journal of Sociology* 108 (5):976–1017.

Morland, K., S. Wing, A. D. Roux, and C. Poole. 2002. Neighborhood characteristics associated with the location of food stores and food service places. *American Journal of Preventive Medicine* 22 (1):23–9.

Paisley, J. W., D. M. Blei, and M. I. Jordan. 2014. Bayesian nonnegative matrix factorization with stochastic variational inference. In *Handbook of mixed membership models and their applications*, ed. E. M. Airoldi, D. Blei, E. A. Erosheva and S. E. Fienberg, 203–22. Boca Raton, FL: CRC Press.

Perchoux, C., B. Chaix, S. Cummins, and Y. Kestens. 2013. Conceptualization and measurement of environmental exposure in epidemiology: Accounting for activity space related to daily mobility. *Health & Place* 21 :86–93.

Phan, X.-H., L.-M. Nguyen, and S. Horiguchi. 2008. Learning to classify short and sparse text & web with hidden topics from large-scale data collections. In *Proceedings of the 17th international conference on World Wide Web*, 91–100. New York, NY: ACM Press.

Pyatt, G. 1976. On the interpretation and disaggregation of Gini coefficients. *The Economic Journal* 86 (342):243–55.

Ren, F. 2016. Activity space. In *Oxford bibliographies in geography*, ed. B. Warf. Oxford University Press. [Online; last modified 28-June-2016] https://doi.org/10.1093/obo/9780199874002-0137.

Sampson, R. J. 2012. *Great American city: Chicago and the enduring neighborhood effect.* Chicago, IL: The University of Chicago Press.

———. 2019. Neighbourhood effects and beyond: Explaining the paradoxes of inequality in the changing American metropolis. *Urban Studies* 56 (1):3–32.

Sampson, R. J., J. D. Morenoff, and F. Earls. 1999. Beyond social capital: Spatial dynamics of collective efficacy for children. *American Sociological Review* 64 (5):633–60.

Sampson, R. J., S. W. Raudenbush, and F. Earls. 1997. Neighborhoods and violent crime: A multilevel study of collective efficacy. *Science* 277 (5328):918–24.

Sastry, N., B. Ghosh-Dastidar, J. Adams, and A. R. Pebley. 2006. The design of a multilevel survey of children, families, and communities: The Los Angeles Family and Neighborhood Survey. *Social Science Research* 35 (4):1000–24.

Sherman, J. E., J. Spencer, J. S. Preisser, W. M. Gesler, and T. A. Arcury. 2005. A suite of methods for representing activity space in a healthcare accessibility study. *International Journal of Health Geographics*, 4:24.

Small, M. L. 2006. Neighborhood institutions as resource brokers: Childcare centers, interorganizational ties, and resource access among the poor. *Social Problems* 53 (2):274–92.

Small, M. L., and M. McDermott. 2006. The presence of organizational resources in poor urban neighborhoods: An analysis of average and contextual effects. *Social Forces* 84 (3):1697–724.

Tan, A.-H. 1999. Text mining: The state of the art and the challenges. In *Proceedings of the PAKDD 1999 workshop on knowledge discovery from advanced databases*, vol. 8, 65–70.

Vallée, J., E. Cadot, F. Grillo, I. Parizot, and P. Chauvin. 2010. The combined effects of activity space and neighbourhood of residence on participation in preventive health-care activities: The case of cervical screening in the Paris metropolitan area (France). *Health & Place* 16 (5):838–52.

Vallée, J., E. Cadot, C. Roustit, I. Parizot, and P. Chauvin. 2011. The role of daily mobility in mental health inequalities: The interactive influence of activity space and neighbourhood of residence on depression. *Social Science & Medicine* 73 (8):1133–44.

Wang, Q., N. E. Phillips, M. L. Small, and R. J. Sampson. 2018. Urban mobility and neighborhood isolation in America's 50 largest cities. *Proceedings of the National Academy of Sciences* 115 (30):7735-7740.

Wasserman, S., and K. Faust. 1994. *Social network analysis: Methods and applications*. Vol. 8 of *Structural analysis in the social sciences*. Cambridge: Cambridge University Press.

Wilson, W. J. 1987. *The truly disadvantaged: The inner city, the underclass, and public policy*. Chicago, IL: The University of Chicago Press.

———. 1996. *When work disappears: The world of the new urban poor*. New York, NY: Knopf.

Wong, D. W. and S.-L. Shaw. 2011. Measuring segregation: An activity space approach. *Journal of Geographical Systems* 13 (2):127–45.

Zenk, S. N., A. J. Schulz, S. A. Matthews, A. Odoms-Young, J. Wilbur, L. Wegrzyn, K. Gibbs, C. Braunschweig, and C. Stokes. 2011. Activity space environment and dietary and physical activity behaviors: A pilot study. *Health & Place* 17 (5):1150–61.

# Appendix 1: The Top Ten Most Visited Block Groups

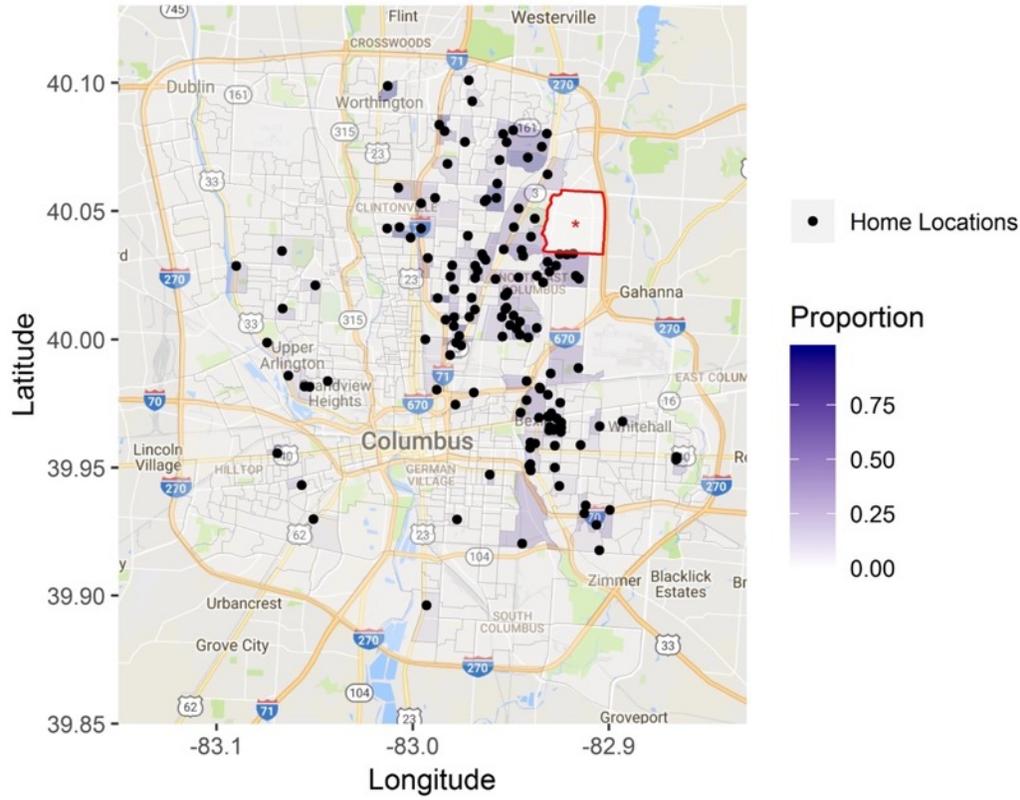

First most visited block group

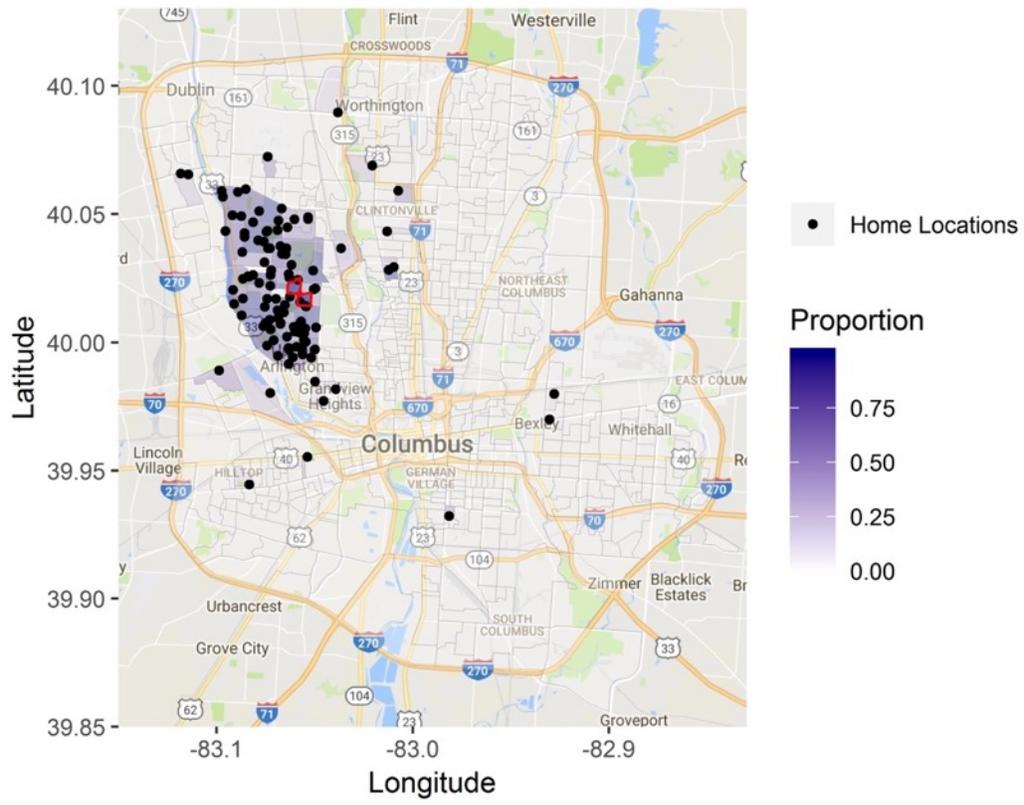

Second most visited block group

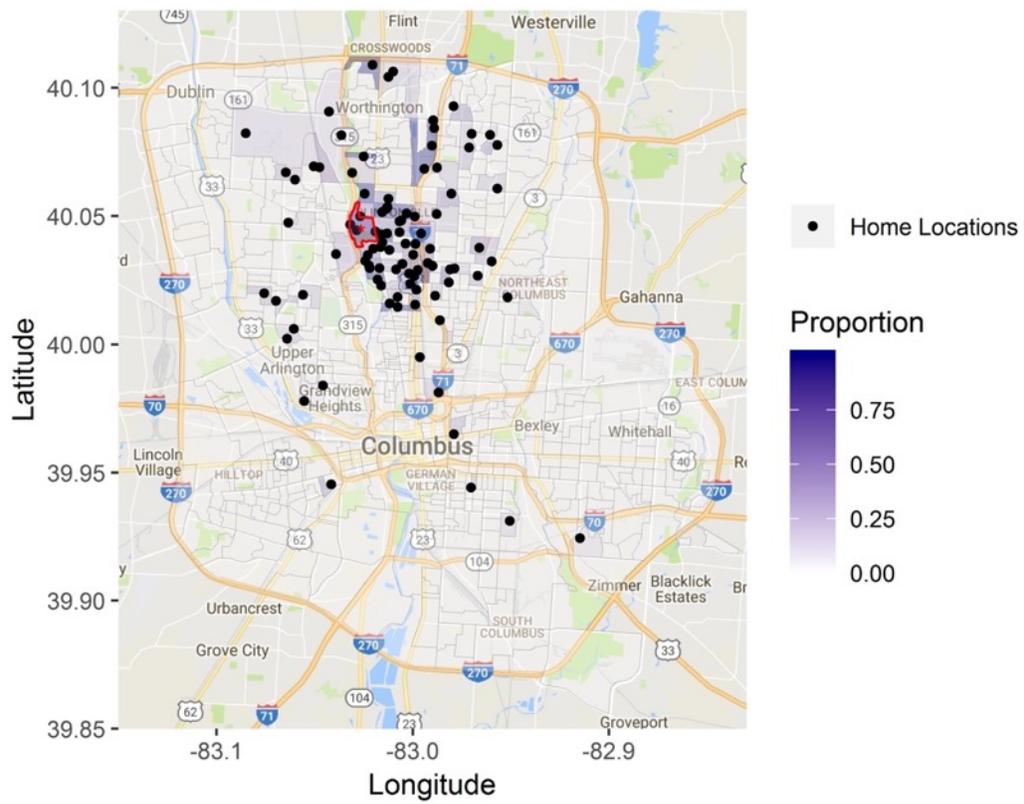

Third most visited block group

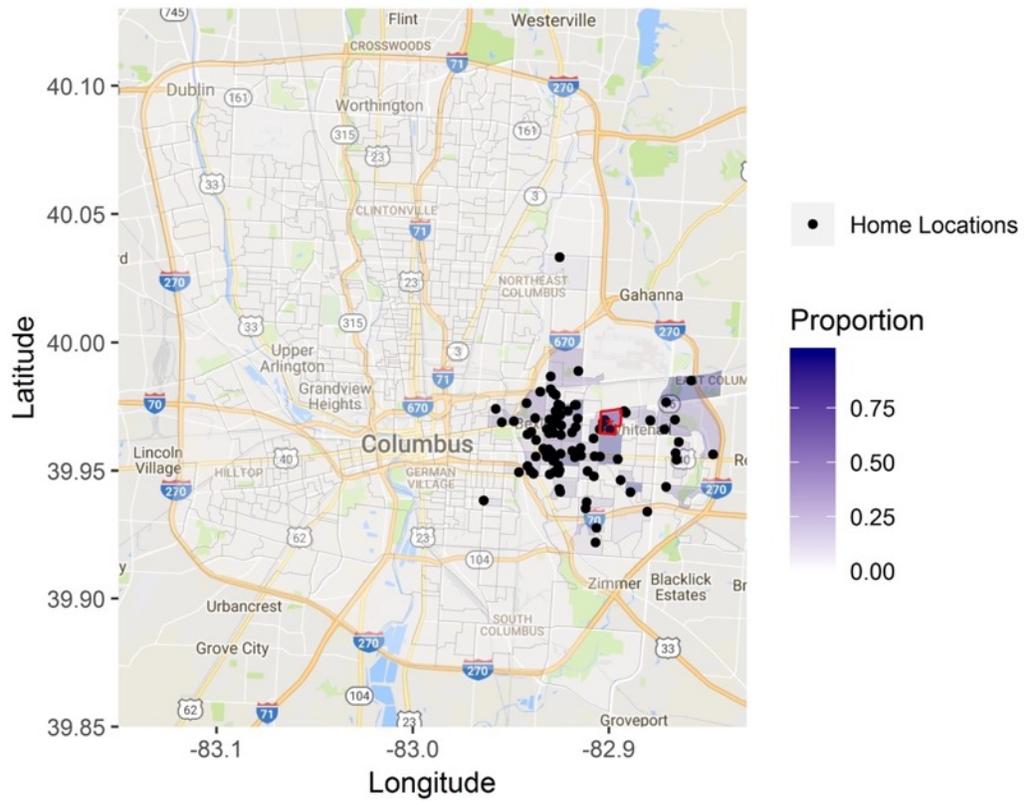

Fourth most visited block group

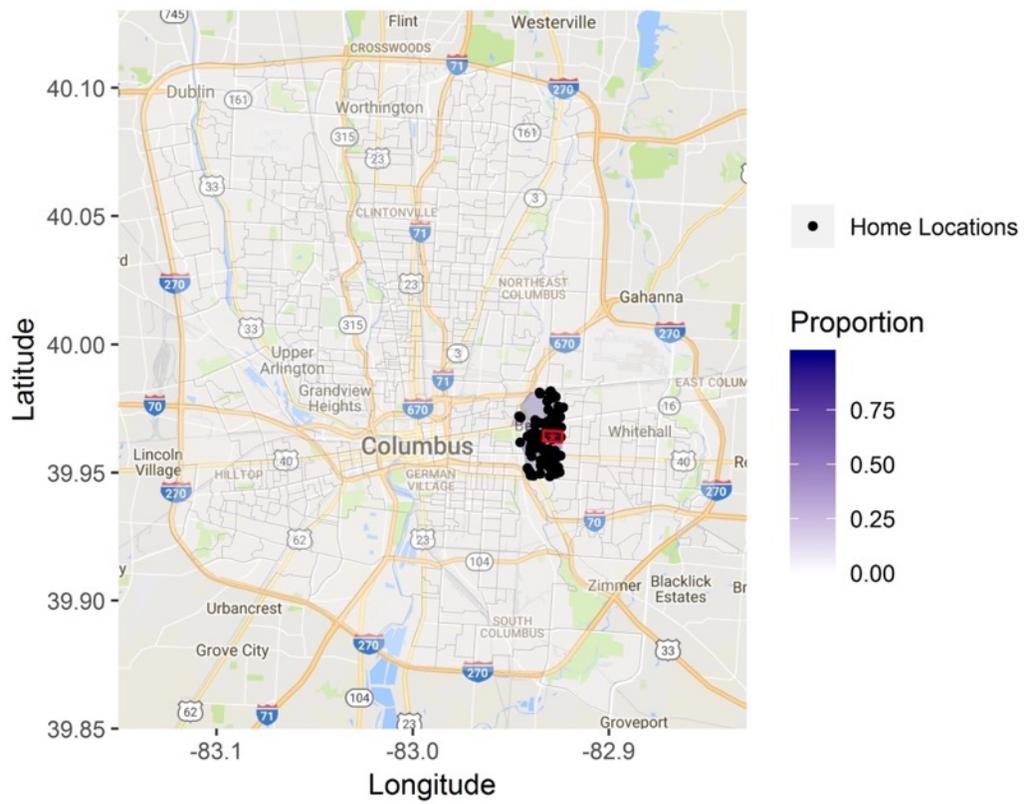

Fifth most visited block group

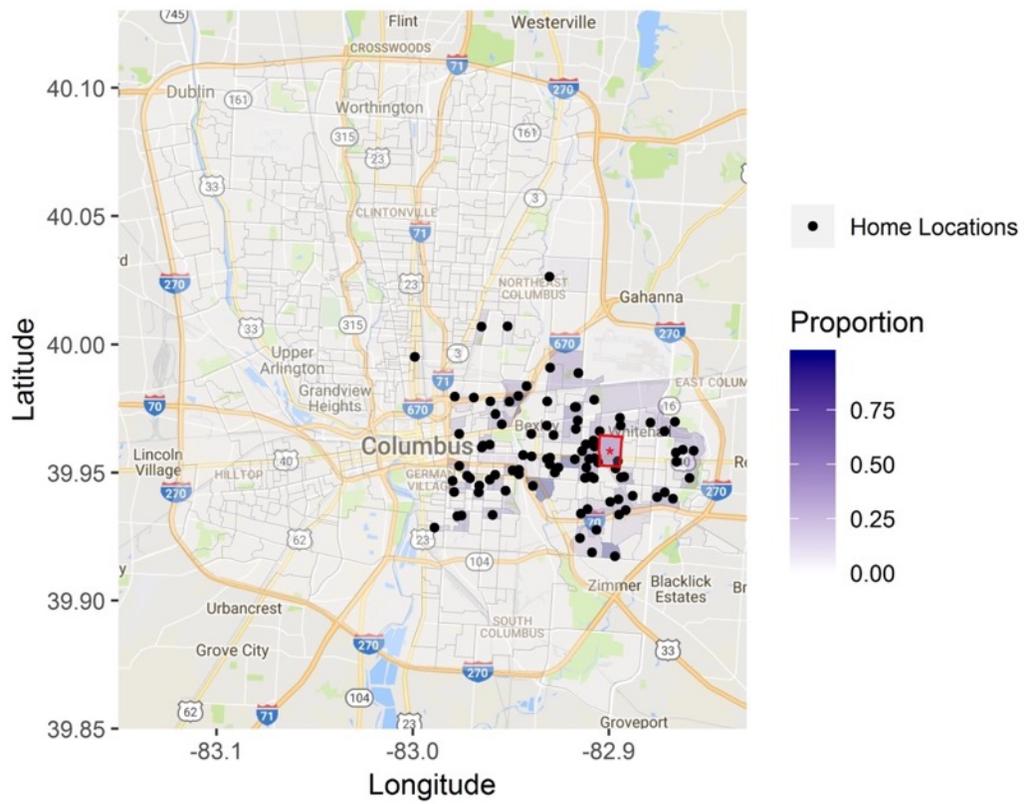

Sixth most visited block group

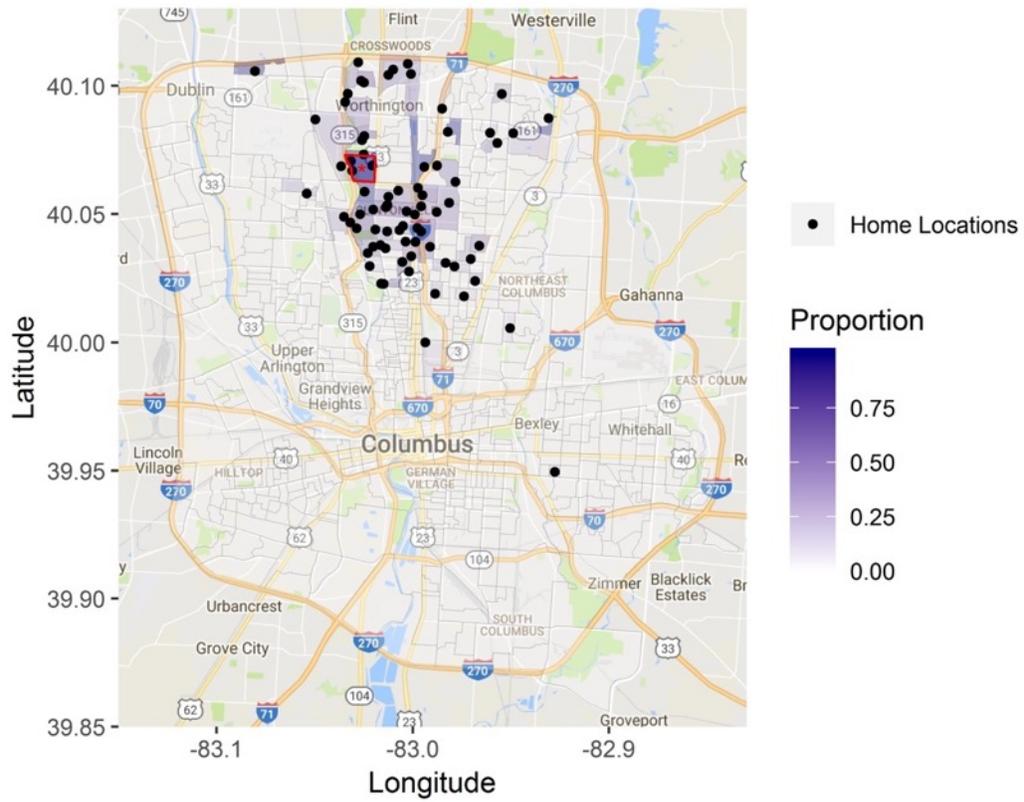

Seventh most visited block group

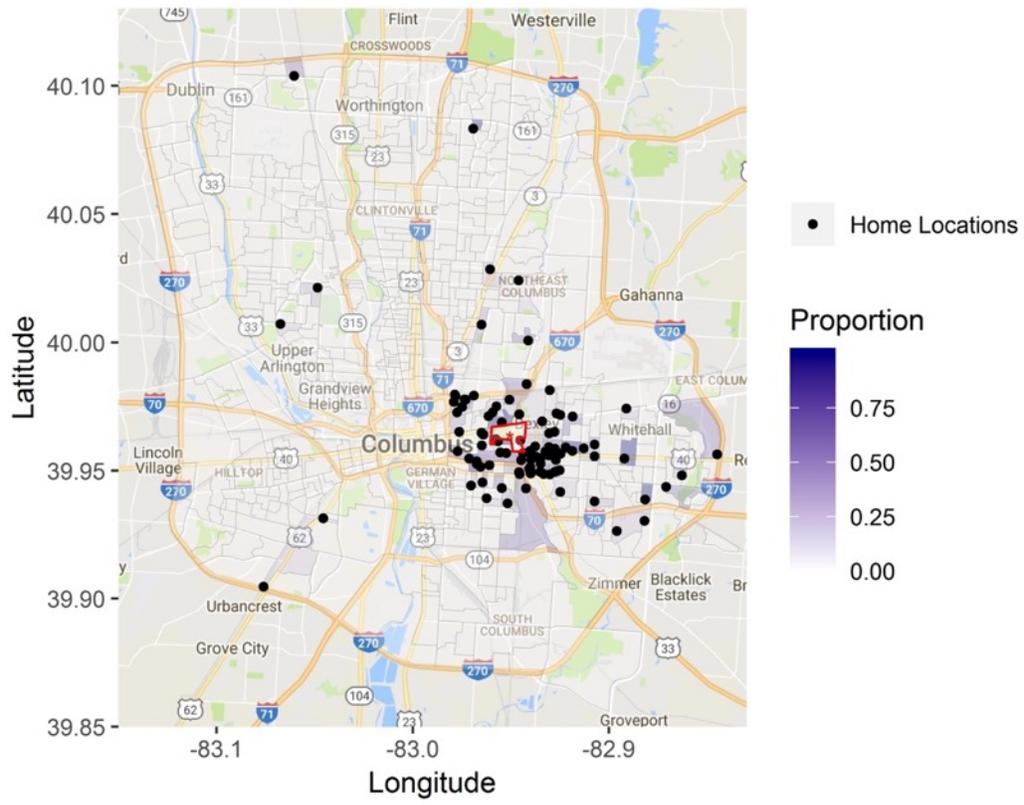

Eighth most visited block group

Ninth most visited block group

Tenth most visited block group

# Appendix 2: The Eighteen Communities Detected from LDA

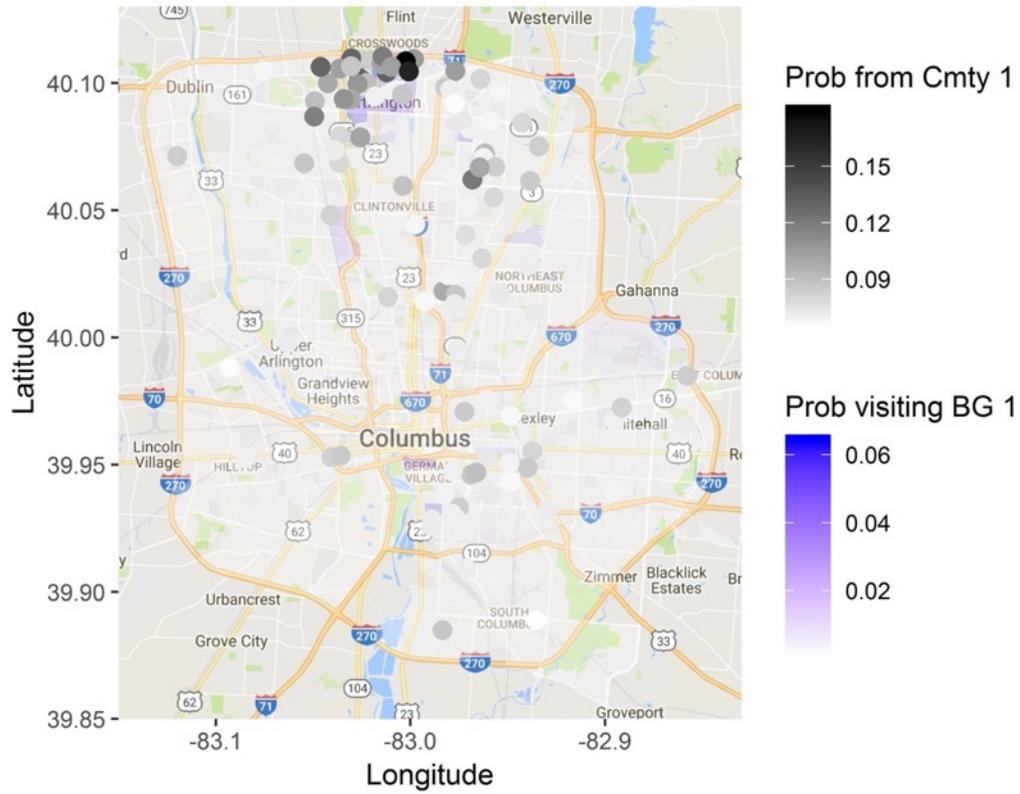

Home locations of community 1

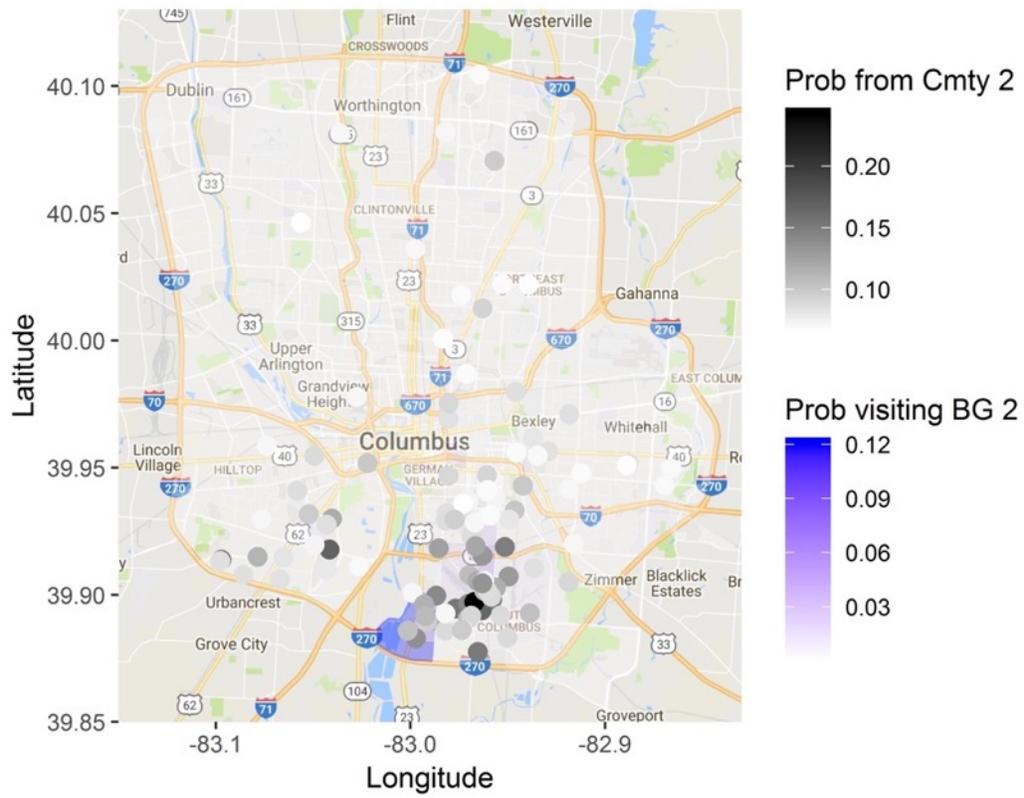

Home locations of community 2

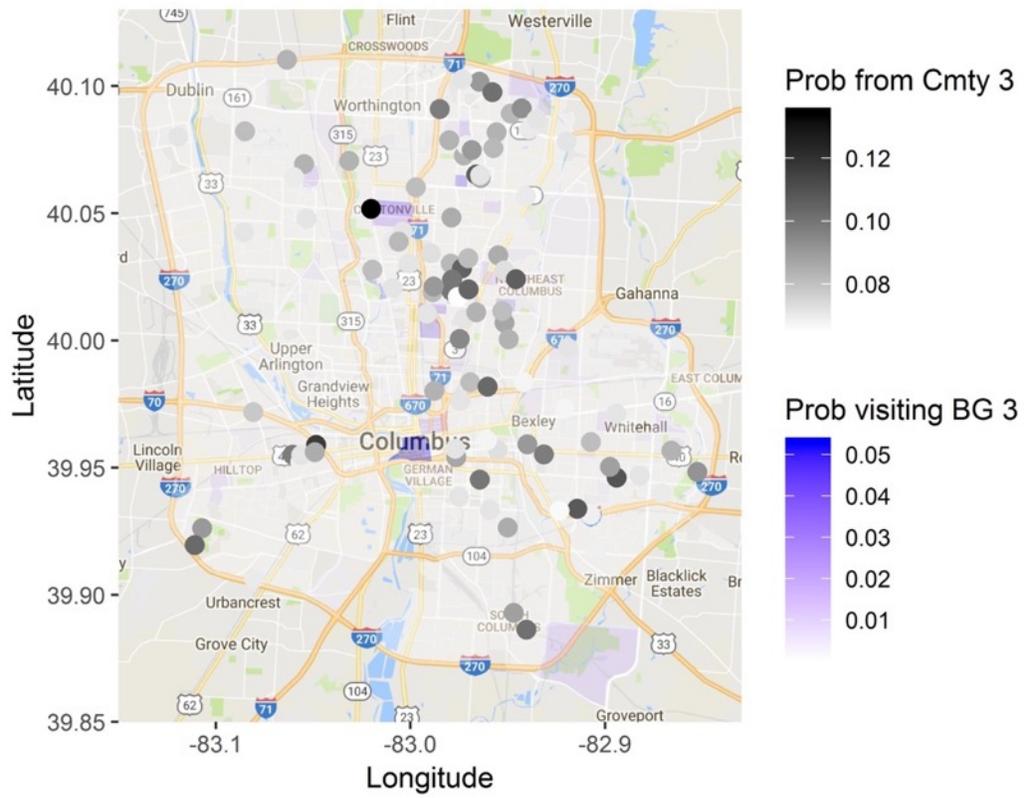

Home locations of community 3

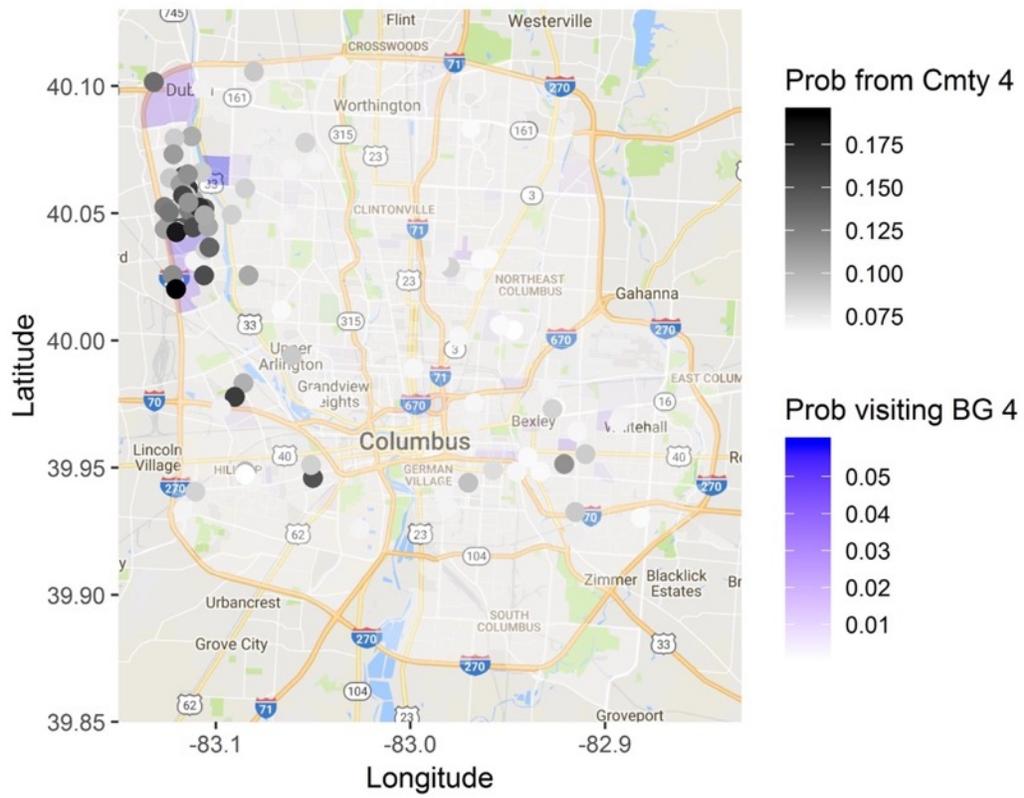

Home locations of community 4

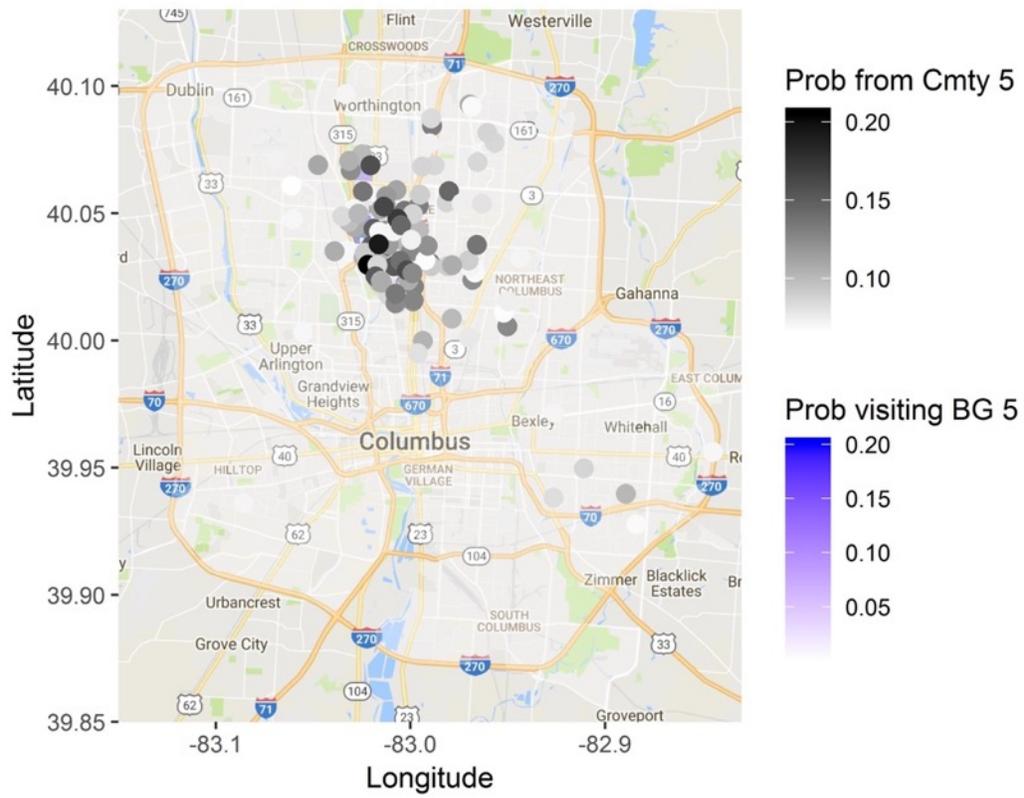

Home locations of community 5

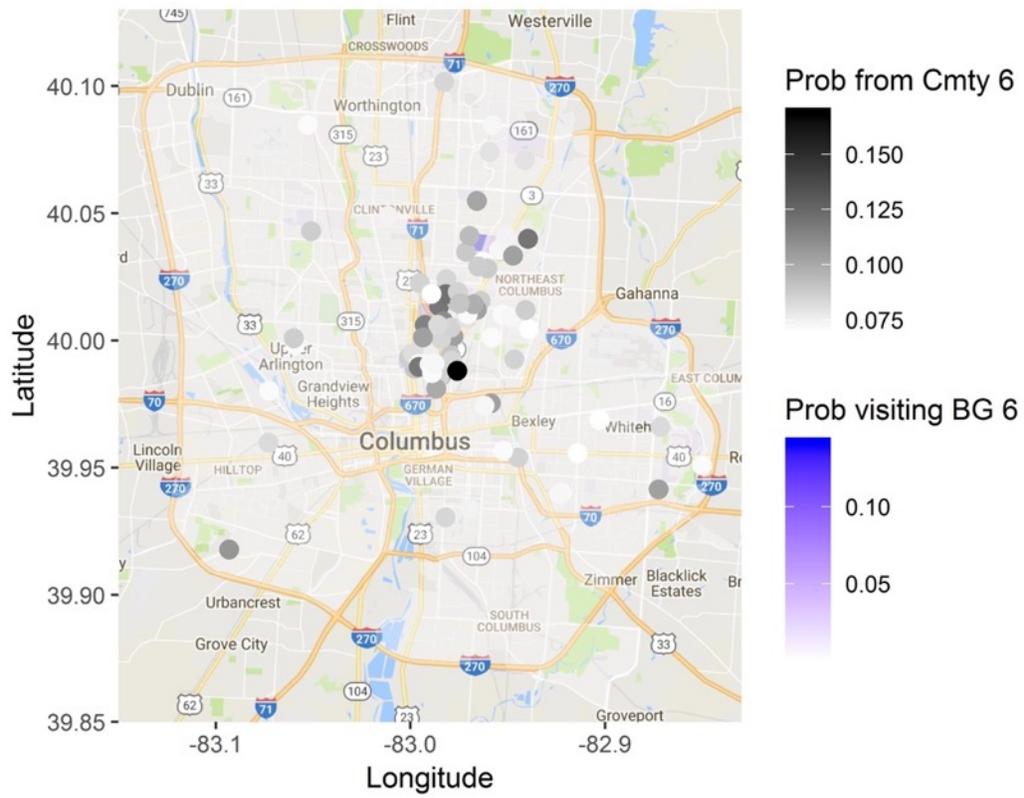

Home locations of community 6

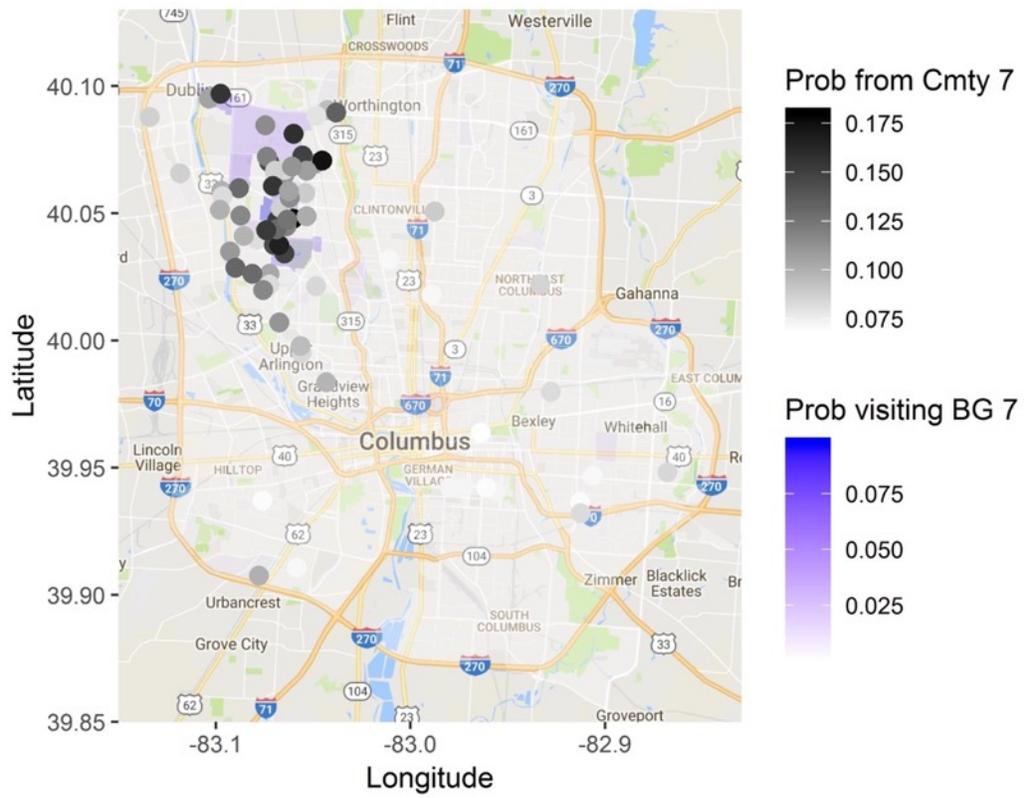

Home locations of community 7

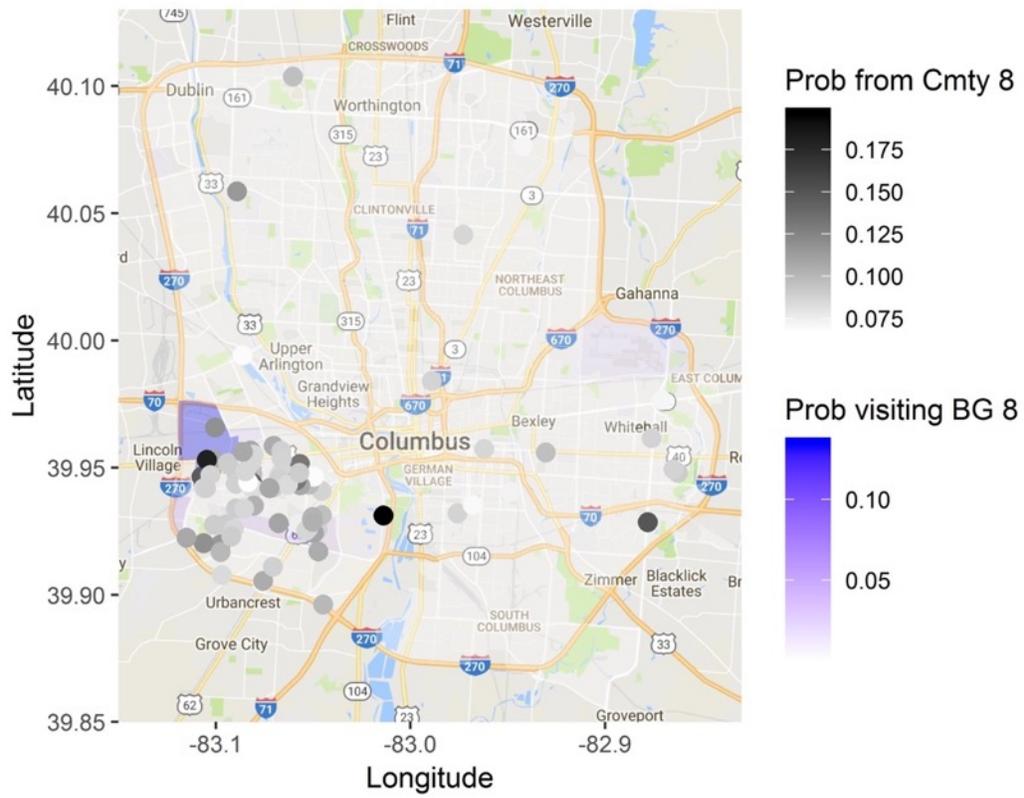

Home locations of community 8

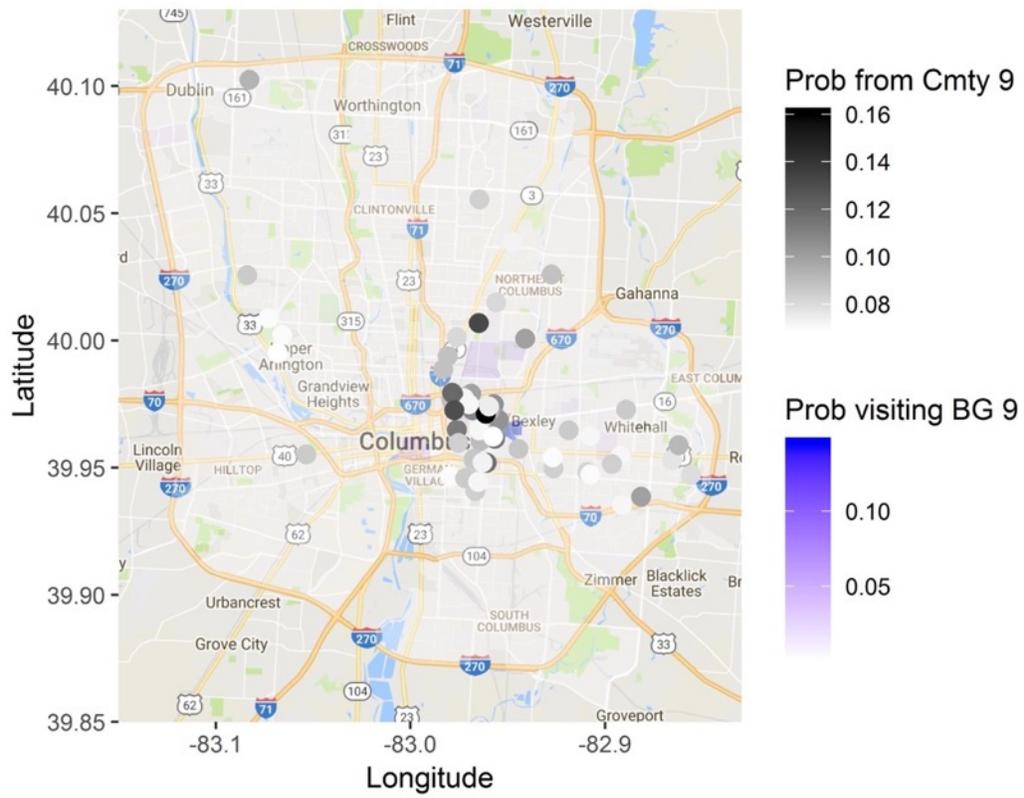

Home locations of community 9

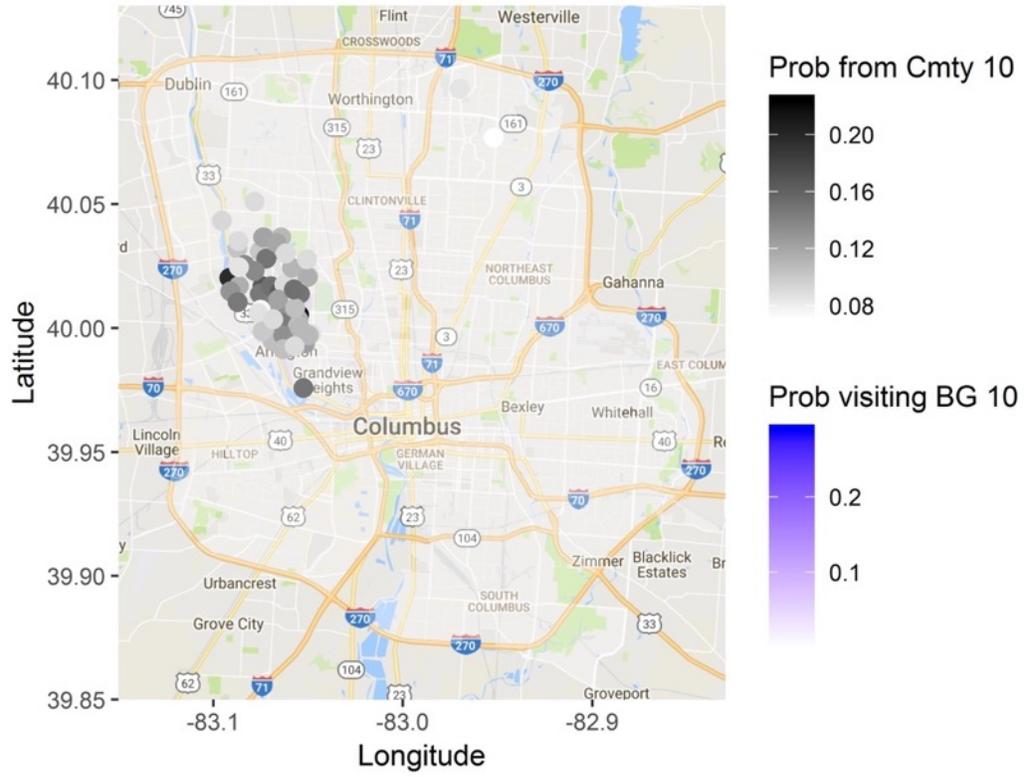

Home locations of community 10

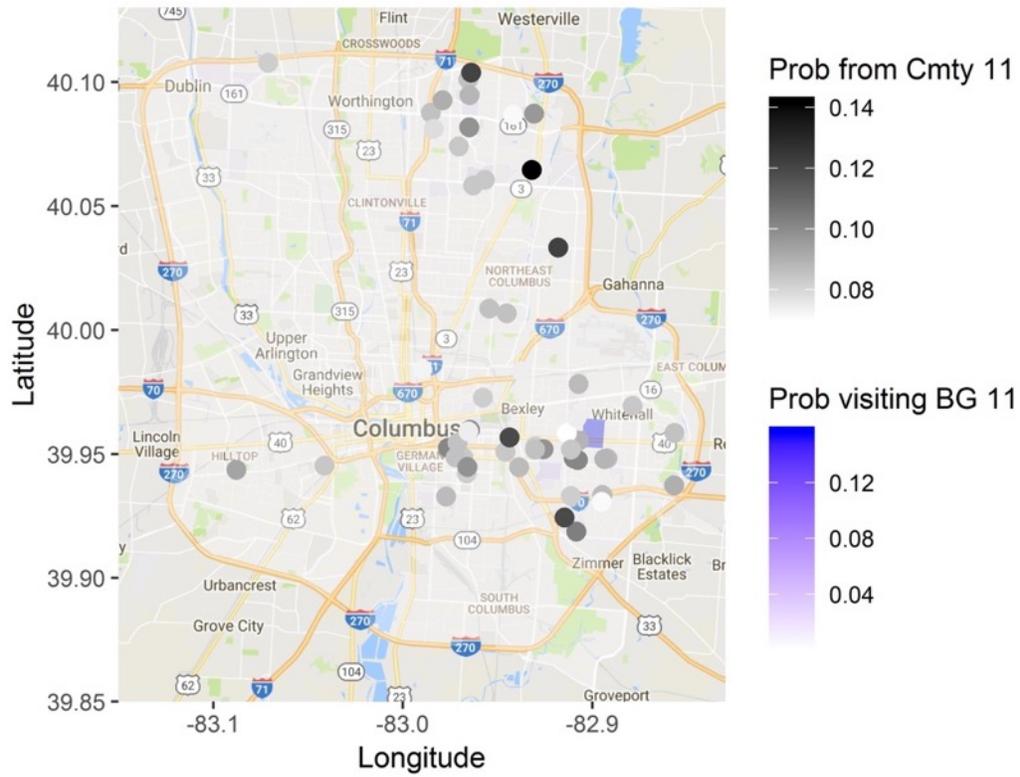

Home locations of community 11

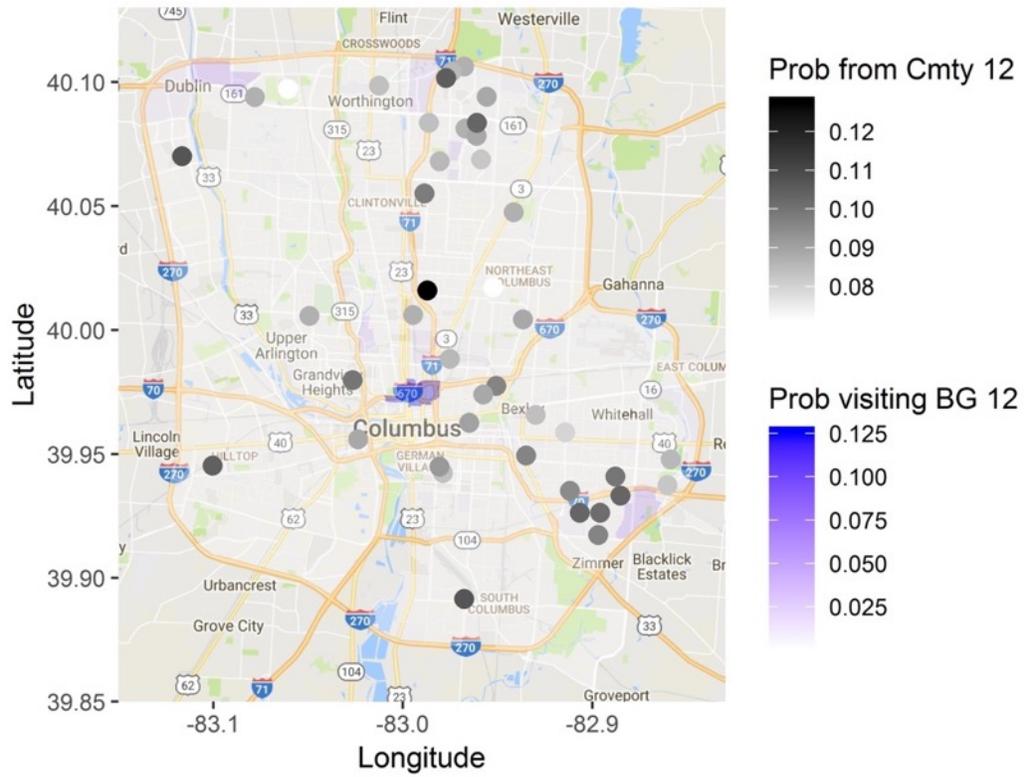

Home locations of community 12

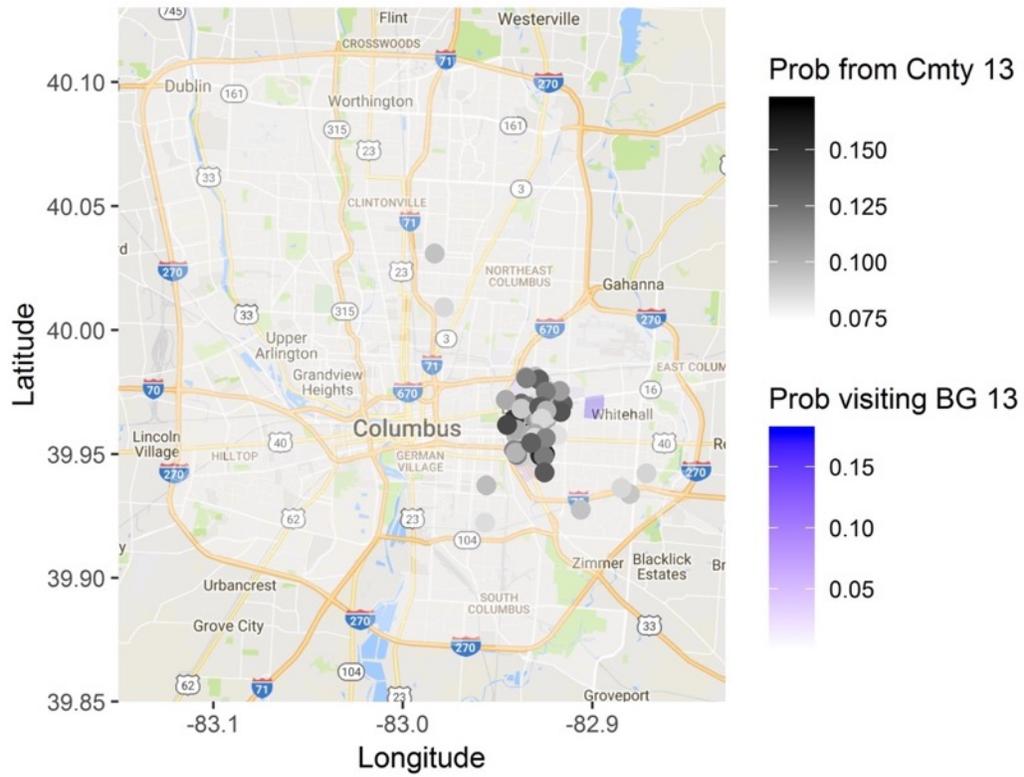

Home locations of community 13

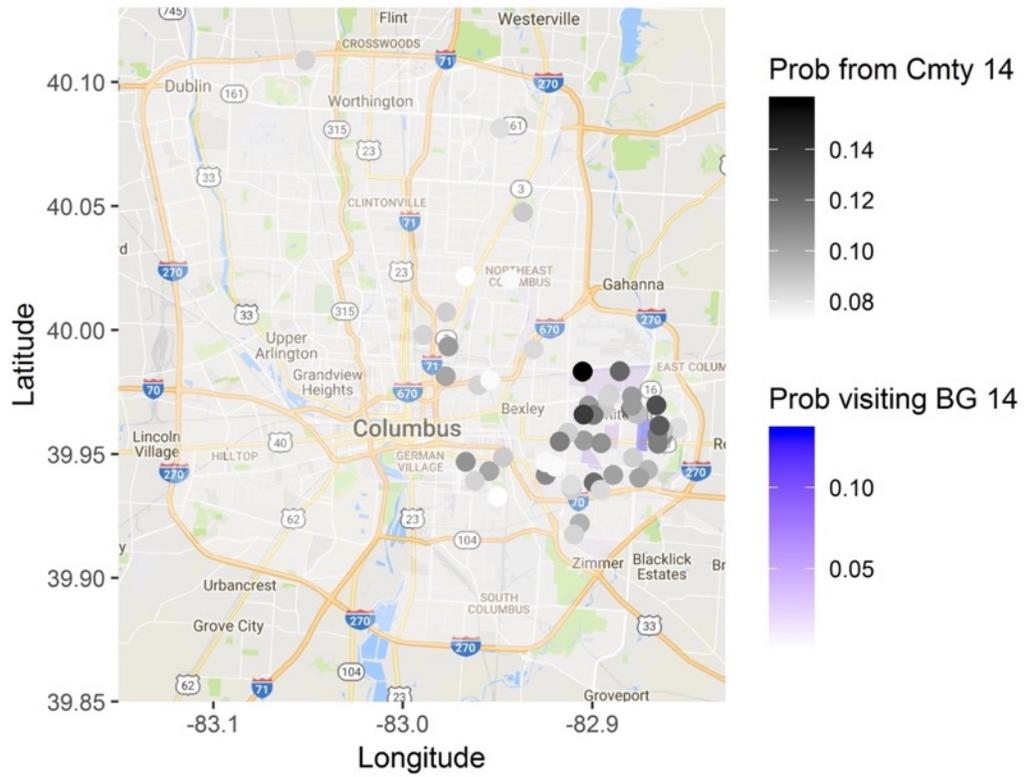

Home locations of community 14

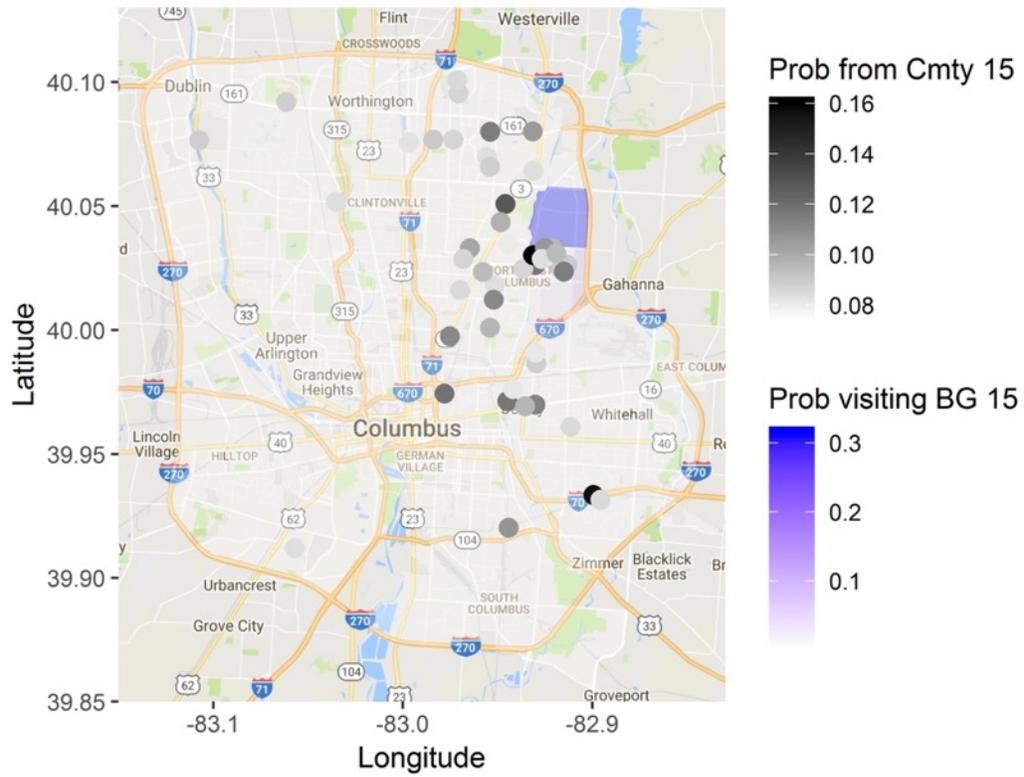

Home locations of community 15

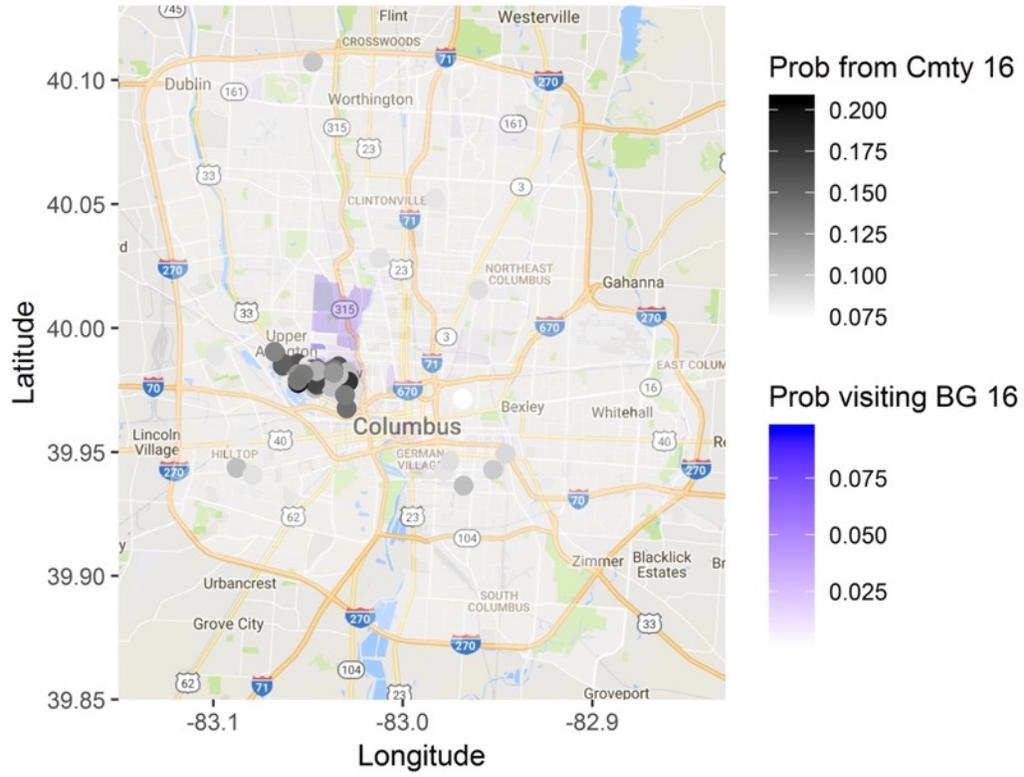

Home locations of community 16

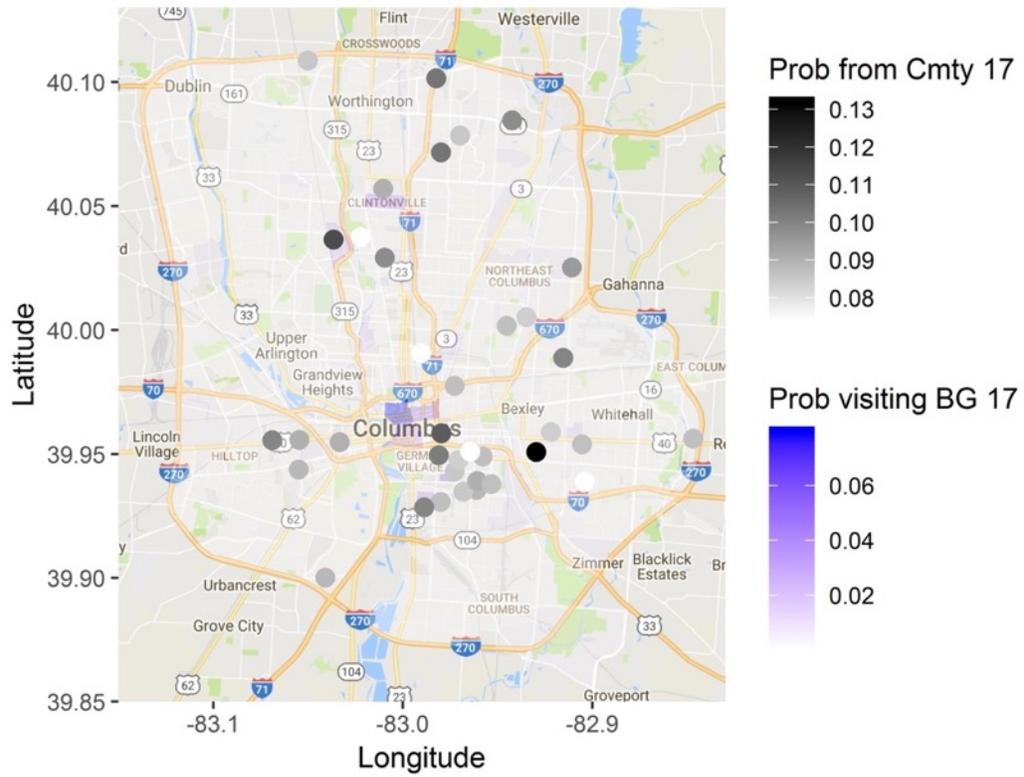

Home locations of community 17

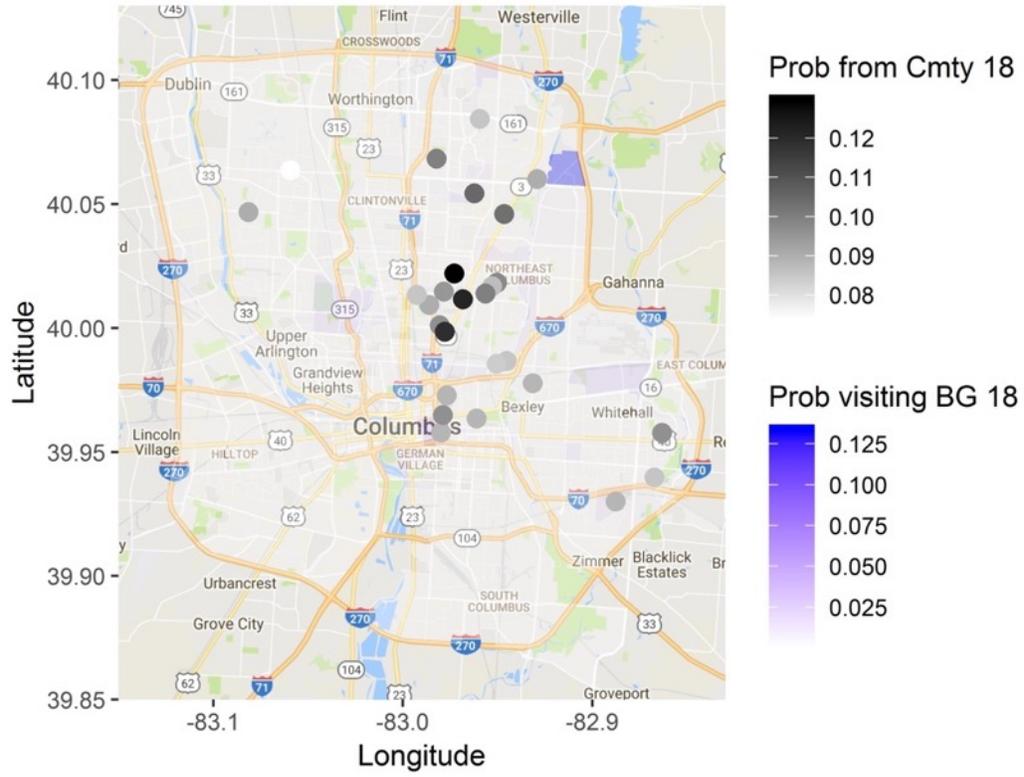

Home locations of community 18

**Tables**

Table 1: Number of caregivers in each community

| Community | 1 | 2 | 3 | 4 | 5 | 6 | 7 | 8 | 9 |
|---|---|---|---|---|---|---|---|---|---|
| Size | 118 | 111 | 98 | 91 | 115 | 83 | 74 | 81 | 67 |
| Community | 10 | 11 | 12 | 13 | 14 | 15 | 16 | 17 | 18 |
| Size | 65 | 54 | 43 | 86 | 56 | 47 | 54 | 37 | 27 |

Table 2: Probability of caregivers from the same neighborhood sharing the same community

| Minimum | 1st Quantile | Median | Mean | 3rd Quantile | Maximum |
|---------|--------------|--------|-------|--------------|---------|
| 0.000   | 0.096        | 0.192  | 0.287 | 0.453        | 1.000   |

Table 3: Number of communities in each neighborhood

| Minimum | 1st Quantile | Median | Mean | 3rd Quantile | Maximum |
|---|---|---|---|---|---|
| 1 | 3 | 4 | 4.594 | 6 | 10 |

Table 4: Mean Gini on concentrated disadvantage, controlling for the mean number of locations provided and the number of caregivers[1]

| Variable | Estimate | SE | p-value |
|---|---|---|---|
| Intercept | 0.000 | 0.032 | 1.000 |
| Concentrated disadvantage | -0.112 | 0.034 | 0.001 |
| Mean number of locations provided | 0.879 | 0.034 | < 0.001 |
| Number of caregivers | 0.055 | 0.032 | 0.087 |

[1] All variables here are standardized to mean = 0 and SD = 1.

Table 5: Mean Gini coefficient on percent black, controlling for the mean number of locations provided and the number of caregivers[1]

| Variable | Estimate | SE | p-value |
|---|---|---|---|
| Intercept | 0.000 | 0.029 | 1.000 |
| Percent black | -0.176 | 0.030 | < 0.001 |
| Mean number of locations provided | 0.881 | 0.030 | < 0.001 |
| Number of caregivers | 0.053 | 0.029 | 0.076 |

[1] All variables here are standardized to mean = 0 and SD = 1.

Table 6: Total variation on mean Gini and percent black, controlling for the mean number of locations provided and the number of caregivers[1]

| Variable | Estimate | SE | $p$-value |
|---|---|---|---|
| Intercept | 0.075 | 0.069 | 0.275 |
| Mean Gini | -0.354 | 0.190 | 0.065 |
| Percent black | 0.153 | 0.074 | 0.040 |
| Mean Gini × percent black | 0.200 | 0.070 | 0.005 |
| Mean number of locations provided | 0.952 | 0.178 | < 0.001 |
| Number of caregivers | -0.164 | 0.065 | 0.013 |

[1] All variables here are standardized to mean = 0 and SD = 1.

# Figure Captions

Figure 1: Distributions of black population in the study area and home and activity locations of caregivers

Figure 1A: Proportion of black at block group level

Figure 1B: Home and activity locations of caregivers

Figure 2: Block group level caregiver visitors

Figure 2A: Number of caregiver visitors of each block group

Figure 2B: The most popular block group and caregivers who visit it on a regular basis

Figure 2C: The 5-th most popular block group and caregivers who visit it on a regular basis

Figure 2D: The 10-th most popular block group and caregivers who visit it on a regular basis

Figure 3: Distribution of home locations by communities

Figure 4: Distributions of household income and race by communities

Figure 4A: Household income

Figure 4B: Race

Figure 5: Distributions of communities 10, 12, 13, and 15

Figure 5A: Community 10

Figure 5B: Community 12

Figure 5C: Community 13

Figure 5D: Community 15

Figure 6: Simulation-based probabilities of having shared locations within and between communities

Figure 7: Communities versus neighborhoods

Figure 7A: Probability of caregivers from the same neighborhood sharing the same community

Figure 7B: Number of communities by number of caregivers sampled from the neighborhood

Figure 8: Total variation on mean Gini and percent black, controlling for the mean number of locations provided and the number of caregivers

# Figures

Figure 1

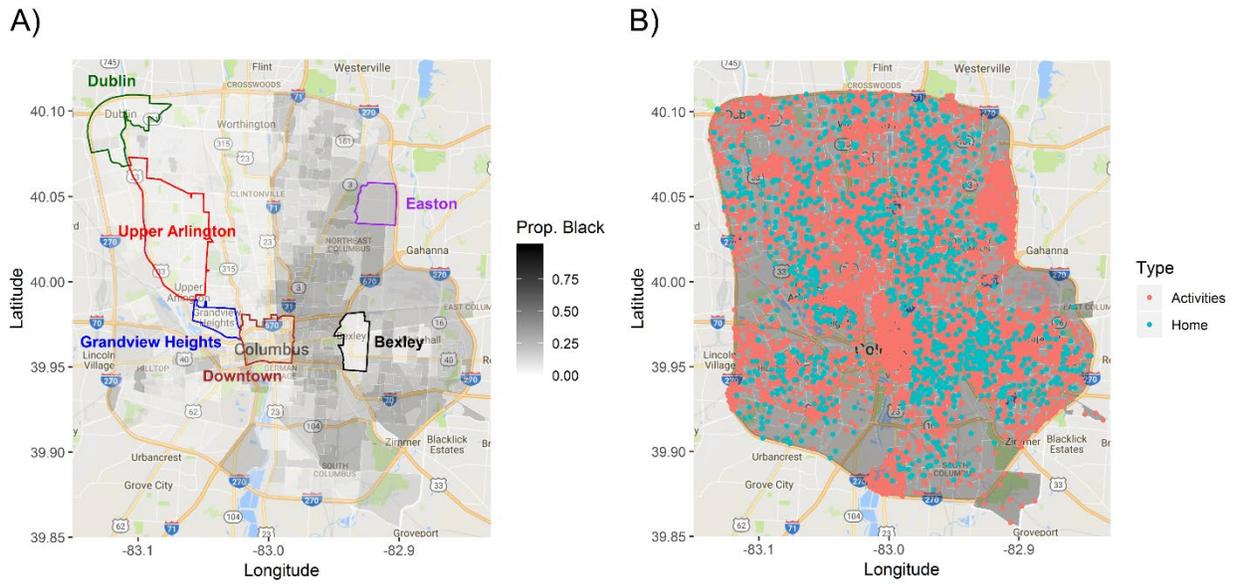

Figure 2

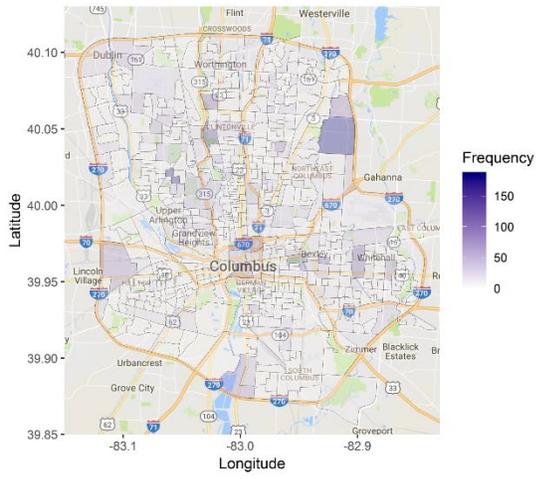
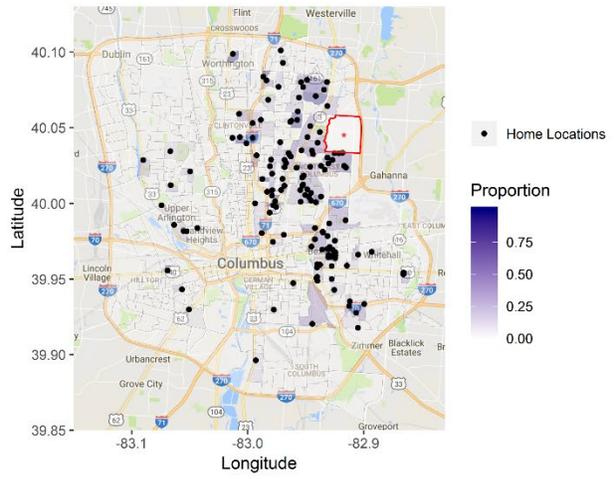

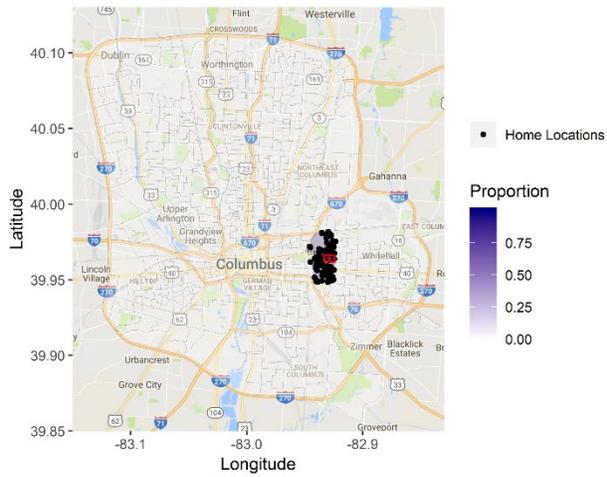
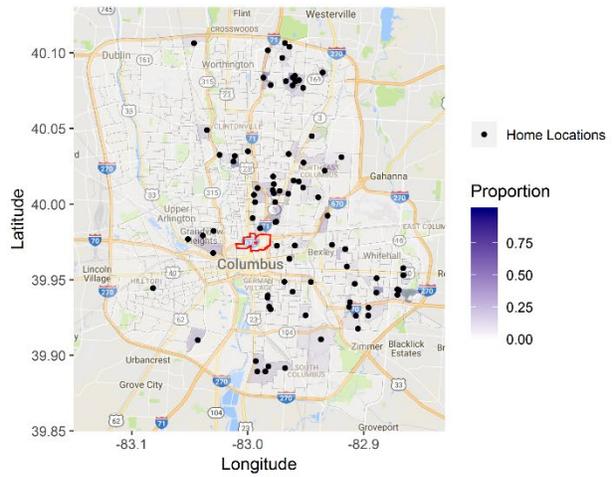

Figure 3

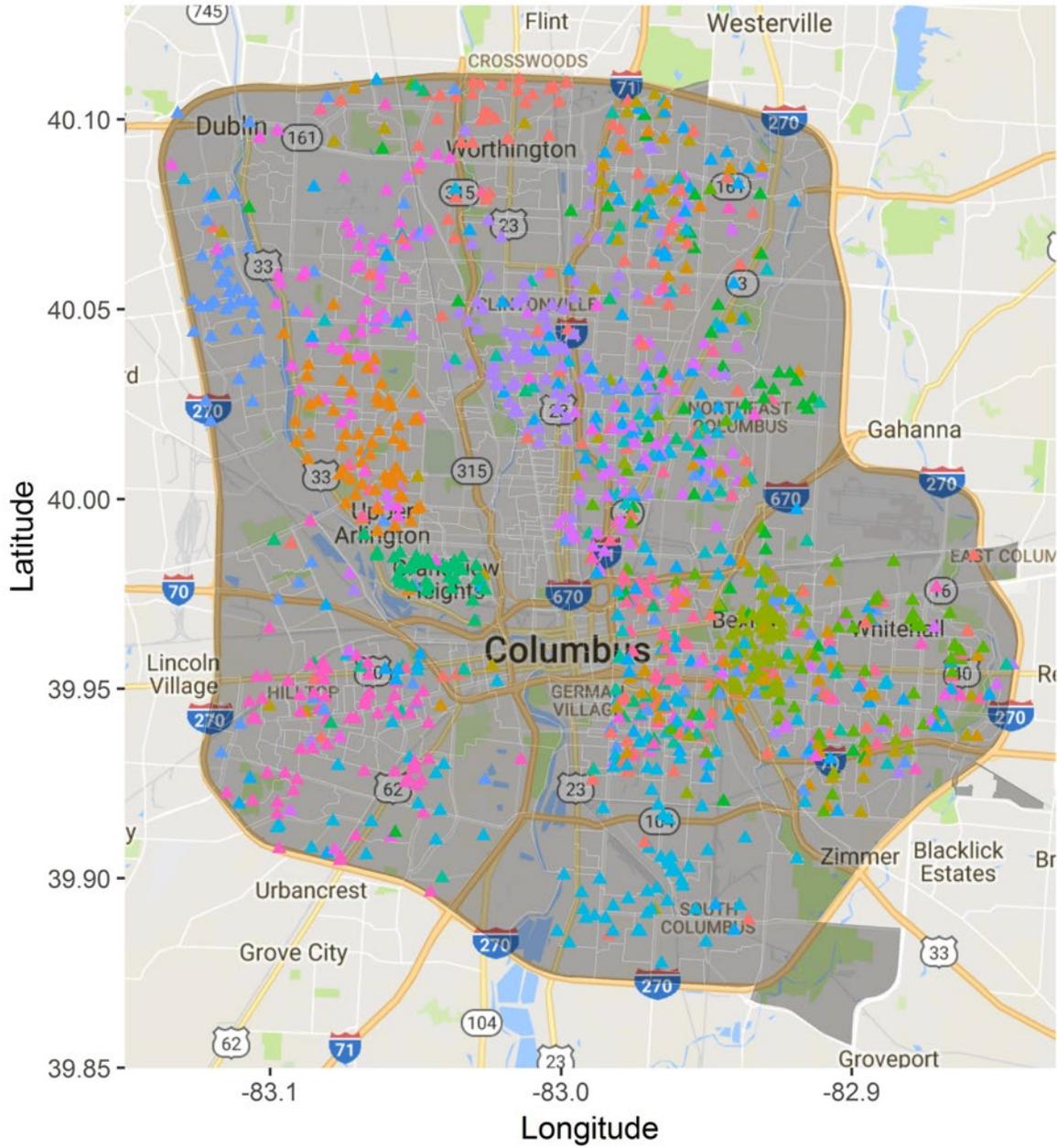

Figure 4

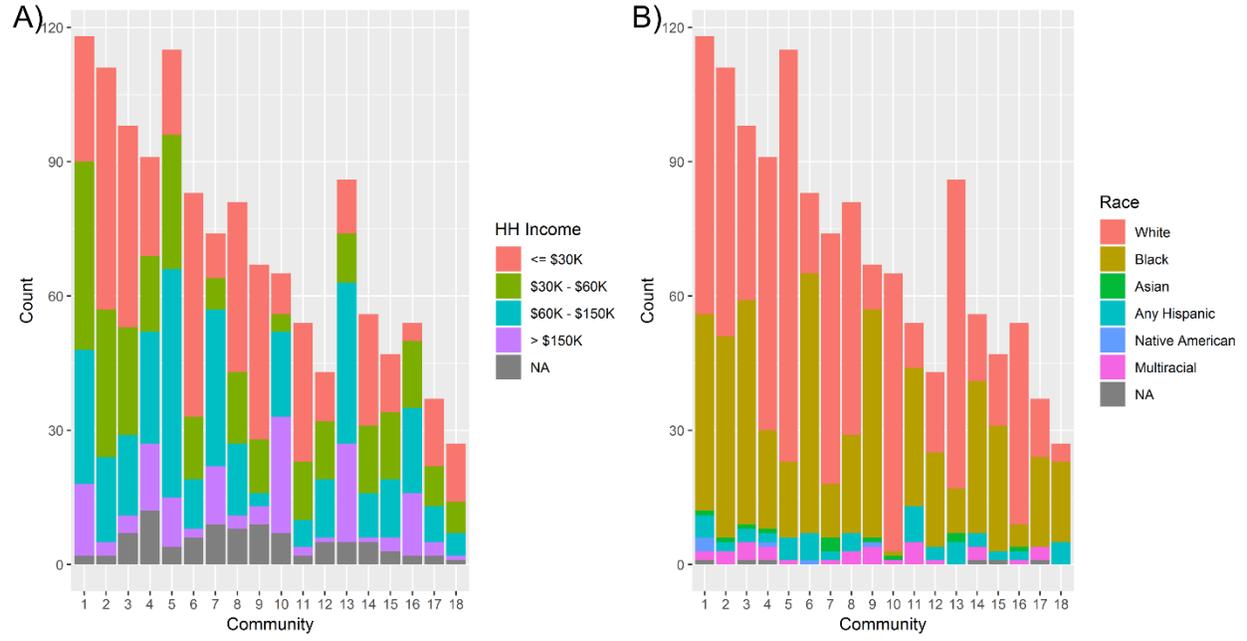

Figure 5

A)
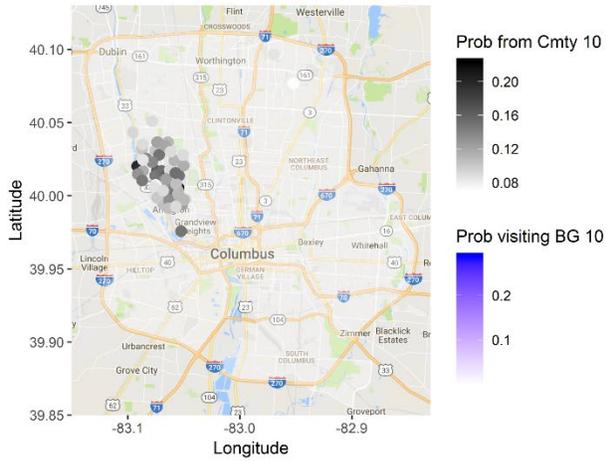

B)
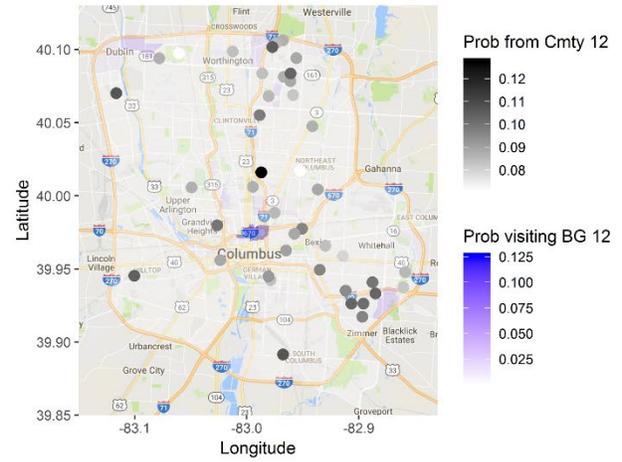

C)
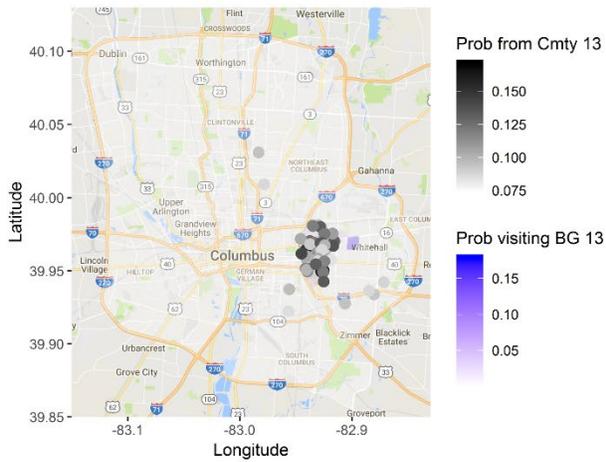

D)
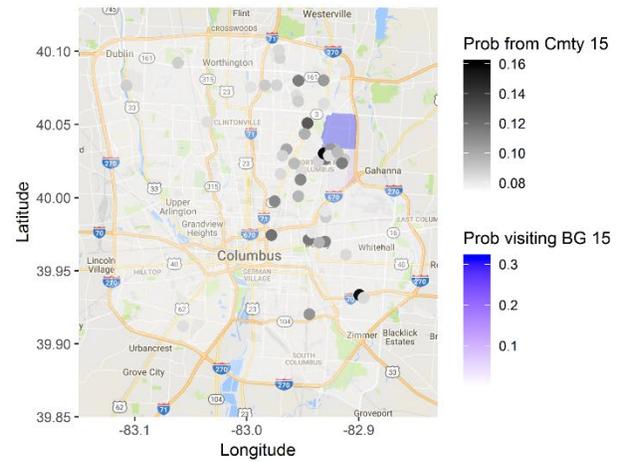

Figure 6

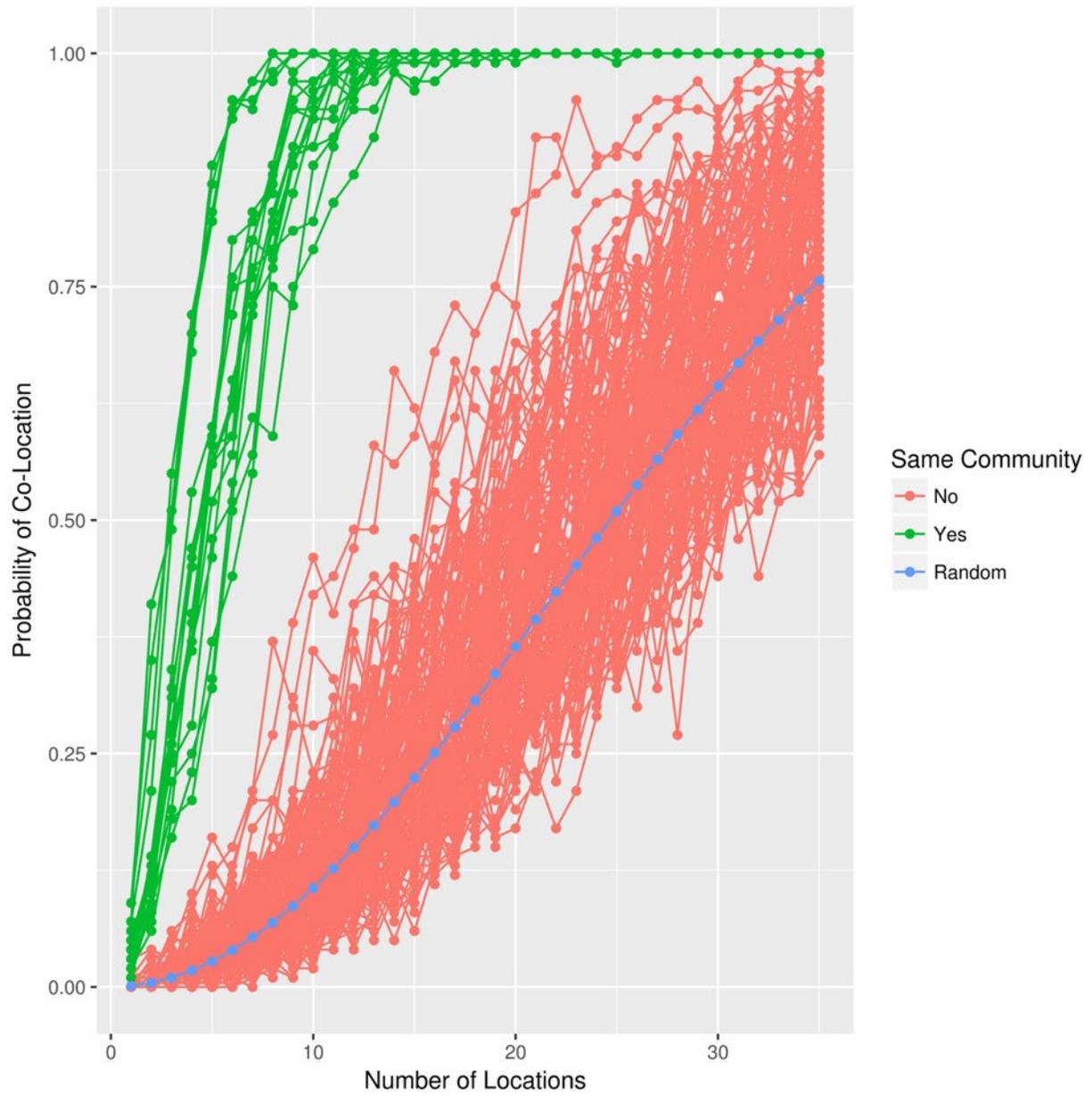

Figure 7

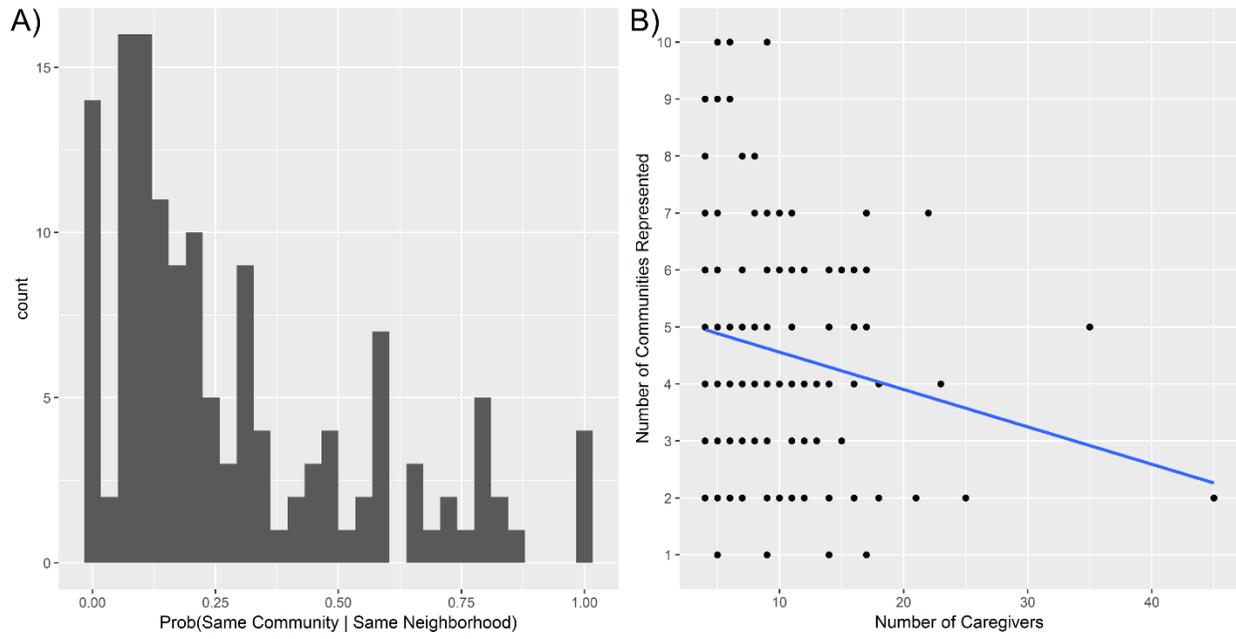

Figure 8

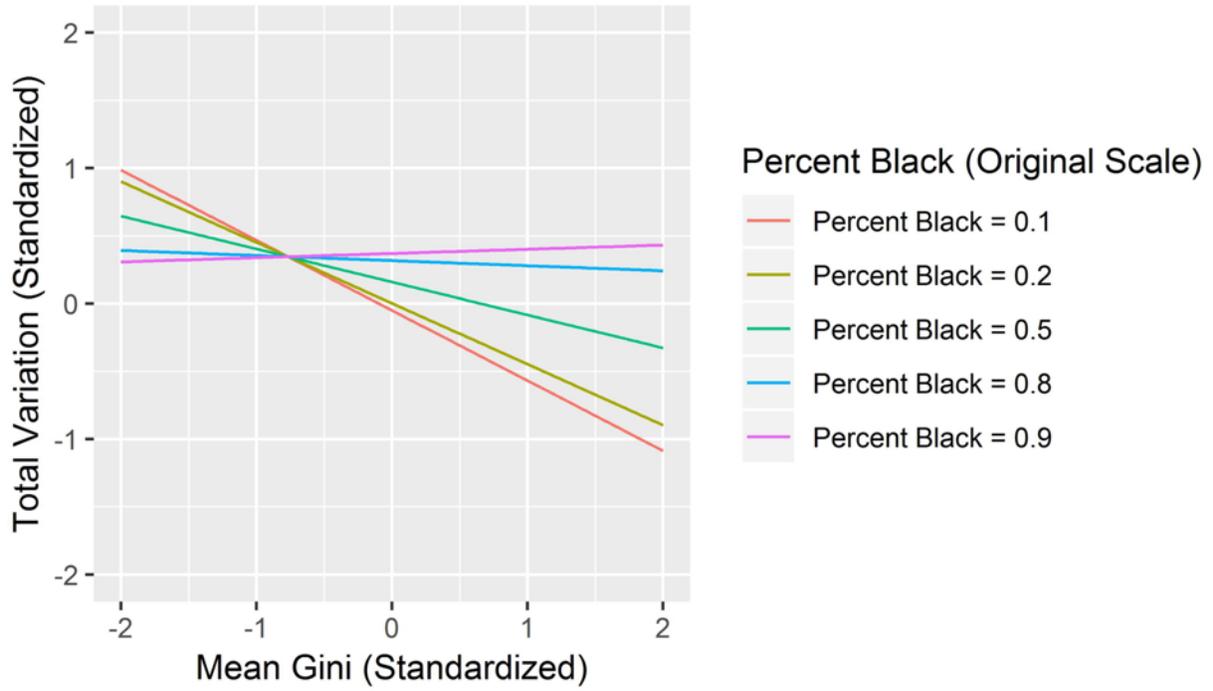